\documentclass[useAMS,usenatbib]{mn2e}
\usepackage{url,times,graphicx,amsmath,amsfonts,amssymb,aas_macros,color,epsfig,varioref,subfigure,comment,natbib}
\usepackage[normalem]{ulem}

\def\bepsilon{\;\pmb{\epsilon}}

\def\bsigma{\;\pmb{\sigma}}

\def\lesssim{{^<_\sim}}
\newcommand{\kms}{{\,\rm km \ s^{-1}}}
\newcommand{\kmsMpc}{{\,\rm km \ s^{-1}Mpc^{-1}}}
\newcommand{\hMpc}{{\ifmmode{h^{-1}{\rm Mpc}}\else{$h^{-1}$Mpc}\fi}}
\newcommand{\hgpc}{{\ifmmode{h^{-1}{\rm Gpc}}\else{$h^{-1}$Gpc}\fi}}
\newcommand{\hmpc}{{\ifmmode{h^{-1}{\rm Mpc}}\else{$h^{-1}$Mpc}\fi}}
\newcommand{\hkpc}{{\ifmmode{h^{-1}{\rm kpc}}\else{$h^{-1}$kpc}\fi}}
\newcommand{\Mpc}{{\ifmmode{{\rm Mpc}}\else{Mpc}\fi}}
\newcommand{\kpc}{{\ifmmode{{\rm kpc}}\else{kpc}\fi}}
\newcommand{\hMsun}{{\ifmmode{h^{-1}{\rm {M_{\odot}}}}\else{$h^{-1}{\rm{M_{\odot}}}$}\fi}}
\newcommand{\hmsun}{{\ifmmode{h^{-1}{\rm {M_{\odot}}}}\else{$h^{-1}{\rm{M_{\odot}}}$}\fi}}
\newcommand{\Msun}{{\ifmmode{{\rm {M_{\odot}}}}\else{${\rm{M_{\odot}}}$}\fi}}
\newcommand{\msun}{{\ifmmode{{\rm {M_{\odot}}}}\else{${\rm{M_{\odot}}}$}\fi}}
\newcommand{\LCDM}{{$\Lambda$CDM}}
\renewcommand{\d}[1]{\ensuremath{\operatorname{d}\!{#1}}}
\def\zcos{ z_{\rm cos} }
\def\zp{ z_{\rm p} }

\def \apjl   {{\it Ap. J. Lett.}}

\def \apjs {{\it Ap. J. Suppl.}}

\def\aap{A\&A}

\def\apjs{ApJS}
\def\apjl{ApJL}

\definecolor{ashgrey}{rgb}{0.7, 0.75, 0.71}
\definecolor{ao(english)}{rgb}{0.0, 0.5, 0.0}

\title
[From Cosmicflows  distance moduli to unbiased  distances and velocities]
{From Cosmicflows  distance moduli to unbiased  distances and peculiar velocities 
}

\author[Y. Hoffman et al.]{
Yehuda Hoffman,$^{1}$\thanks{E-mail: Hoffman@huji.ac.il (YH)}
Adi Nusser,$^{2}$
Aur\'elien Valade,$^{3,4}$
Noam I. Libeskind,$^{3,4}$
R.  Brent Tully$^{5}$
\\
$^{1}$Racah Institute of Physics, Hebrew University, Jerusalem 91904, Israel\\
$^{2}$Physics Dept, The Technion, Haifa 32000, Israel\\
$^{3}$Leibniz Institut f\"ur Astrophysik Potsdam (AIP), An der Sternwarte 16, D-144 Potsdam, Germany\\
$^{4}$University of Lyon, UCB Lyon 1, CNRS/IN2P3, IUF, IP2I Lyon, France\\
$^{5}$Institute for Astronomy, University of Hawaii, Honolulu HI 96822, USA
}

\begin{document}

\date{Submitted XXXX XXX XXXX}

\pagerange{\pageref{firstpage}--\pageref{lastpage}} \pubyear{2019}

\maketitle

\label{firstpage}


\begin{abstract}
Surveys of galaxy  distances  and  radial peculiar velocities can be used to reconstruct the large scale   structure. Other than systematic errors in the zero-point calibration of the galaxy distances the main source  of   uncertainties   of such data are errors on the distance moduli, assumed here to be Gaussian and thus turn  into lognormal errors on distances and velocities. Naively treated,  it  leads to  spurious nearby outflow   and strong infall at larger distances. The lognormal bias is corrected here  and   tested against mock data   extracted from a \LCDM\ simulation, designed to statistically follow the grouped Cosmicflows-3  (CF3) data.  Considering  a subsample of data points, all of  which have the same true distances or same redshifts,  the lognormal bias arises because the means of the distributions of observed distances and velocities are skewed off the means  of the true distances and velocities. Yet, the medians are invariant under the lognormal transformation. That  invariance   allows the Gaussianization of the distances and velocities and the removal of the lognormal bias. 

This  Bias   Gaussianization correction (BGc) algorithm is tested against   mock CF3 catalogs. The test consists of a comparison   of the BGC estimated  with the simulated distances and velocities and of an examination of the Wiener filter reconstruction from the BGc data. Indeed, the BGc eliminates the lognormal bias.

The estimation of Hubble's constant ($H_0$) is also tested. The  residual of the BGc estimated $H_0$ from the simulated  values  is $-0.6\pm0.7\kmsMpc$, and is dominated by the cosmic variance. The  BGc correction of the actual CF3 data yields $H_0=75.8\pm1.1\kmsMpc$.

\end{abstract}

\section{Introduction}

In the standard model of cosmology departures from uniform density and pure Hubble flow are strongly coupled - density irregularities induce peculiar velocities on top of the Hubble flow; peculiar  velocities   drive  the matter away from uniform distribution. The equation of continuity tells it all  \citep{1980lssu.book.....P,2008cosm.book.....W}.  This is why surveys of peculiar velocities of galaxies  play such  an important  role in unveiling   the  underlying - luminous and dark - mass  distribution  in  the  nearby universe  
\citep[][is only a partial list]{1986ApJ...307...91L,1988ApJ...326...19L,1990ApJ...364..349D,1996ApJ...468L...5D,2006ApJ...653..861M}. 
Peculiar velocity surveys have also been used to constraints   cosmological parameters 
\citep[e.g.][]{1993ApJ...412....1D,2001MNRAS.326..375Z,2011ApJ...736...93N,2017MNRAS.470..445N,2018MNRAS.481.1368P}.
Velocity surveys are less effective in constraining  the values of cosmological parameters compared with other probes - CMB  anisotropies in particular  \citep[e.g.][]{Planck:2013}, but they are our only means  for directly mapping the (total) mass distribution  in the nearby universe.

Of particular interest   is the Cosmicflows project\footnote{https://www.ip2i.in2p3.fr/projet/cosmicflows} of measuring and compiling distances and redshifts of galaxies,  and thereby estimating their peculiar velocities. Three data catalogs have been  released so far:  Cosmicflows-1   \citep{2008ApJ...676..184T}, Cosmicflows-2  \citep{2013AJ....146...86T} and Cosmicflows-3 \citep[hereafter CF3]{2013AJ....146...86T}. The Constrained Local UniversE Simulation's (CLUES) collaboration\footnote{https://www.clues-project.org/cms/} primary focus  is on   the reconstruction of the present epoch  density  and velocity  fields 
\citep[e.g.][]{2012ApJ...744...43C,2013AJ....146...69C,
2014Natur.513...71T,
2017NatAs...1E..36H,
2017ApJ...845...55P,
2020ApJ...897..133P}
and on setting  initial conditions for constrained simulations of the local universe
\citep[e.g.][]{2008MNRAS.386..390H,
2011MNRAS.417.1434F,
2014NewAR..58....1Y,
2016MNRAS.455.2078S,
2018NatAs...2..680H,
2020MNRAS.496.4087O,2020MNRAS.498.2968L}
from the Cosmicflows data.

Theorists like peculiar velocities  -  their emergence in the standard  cosmological model, the \LCDM\ model, is well understood and in the linear regime the velocity and the density fields are related by a simple linear relation. This stands in sharp contrast to the difficulties arising in estimating  velocities  from observations.  In the rest of the paper we shall focus on one particular data catalog - the CF3 data - and address various  issues  concerning  these data that arise  within the framework of the linear reconstruction of the density  and 3D velocity fields from radial peculiar velocities by means of the Wiener filter and constrained realizations  of Gaussian random fields  \citep[WF/CRs;][]{1991ApJ...380L...5H,1995ApJ...449..446Z,1999ApJ...520..413Z}.

 The CF3 data \citep{2016AJ....152...50T}  is based on the two early releases of the Cosmicflows data \citep{2008ApJ...676..184T,2013AJ....146...86T} augmented by more recently  observed  galaxies observed with  with {\it Spitzer Space Telescope}  for which TF  \citep{1977AA....54..661T} distances are estimated  and by the    addition from the literature of  the extensive Fundamental Plane (FP) sample derived from the Six Degree Field Galaxy Survey (6dFGS) of the southern celestial hemisphere \citep{2014MNRAS.445.2677S}.  \cite{2016AJ....152...50T} describes the different biases corrections, the homogenization  of the different data sources and the zero-point calibration of the data thereby taking care of the sources of the systematic errors. Here we assume that the  systematic uncertainties of the CF3 data have been properly dealt with, leaving us  with the genuine statistical  uncertainties that are being addressed here.

\begin{figure}
\centerline{
\includegraphics[width=1.\columnwidth]{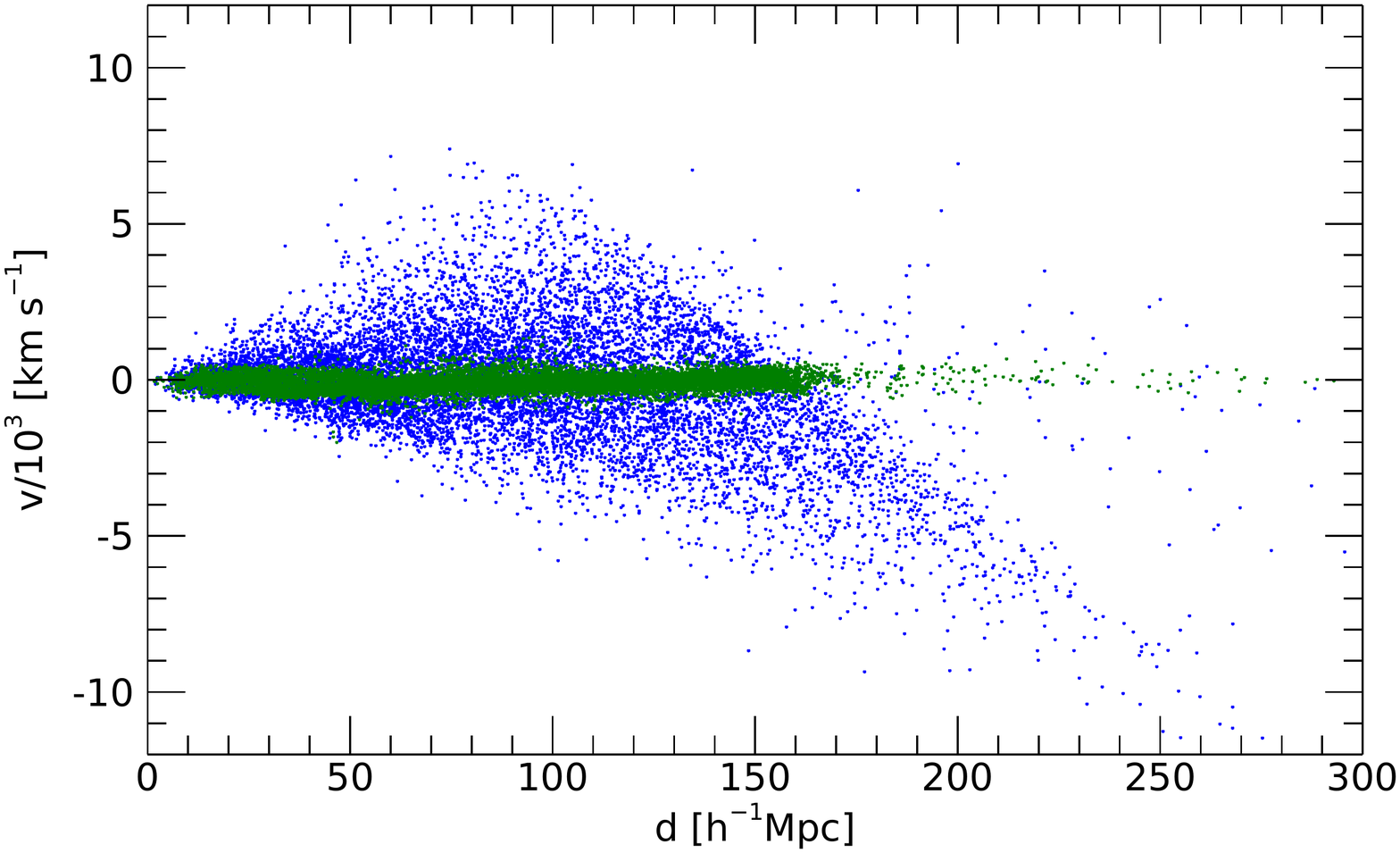} 
}
\caption{ A scatter plot of radial peculiar velocities vs. distance of data points drawn from a CF3-like mock catalog. The blue symbols represent the estimated distances and velocities of the `observed' data points. The actual values, drawn from the simulation, are presented for reference in green. All the different scatter plots shown in the paper are based on data drawn from one of the mock catalogs. }
\label{fig:V-vs-D}
\end{figure}

Peculiar velocities surveys  are notoriously known to suffer from the so-called Malmquist bias \citep[after][]{1920MeLuS..22....3M,1924MeLuS..32....3M}, which has been extended  to encompasses  a variety of biases stemming from the fact that such surveys are essentially flux limited and are therefore subject to selection biases. The various aspects of the Malmquist bias have been classified as the  selection, homogenous and non-homogeneous Malmquist biases
\citep{1988ApJ...326...19L,
1992scma.conf..201L, 
1995PhR...261..271S,
2015MNRAS.450.2644S}. There is another  bias that is often being associated  with the Malmquist bias - one that stems from the Gaussian errors on the distance moduli and thereby results in lognormal uncertainties on the  distances. This lognormal bias skews the distribution of the   observed   velocities at a given  observed   distance shell towards negative values.  Fig. \ref{fig:V-vs-D} manifests very clearly the lognormal bias -  it shows the scatter of radial peculiar velocities vs. distances,  true and `observed' ones, of a mock catalog drawn  from a \LCDM\ simulation (see below).

A brief and incomplete review of mostly recent attempts  to correct for the Malmquist biases in velocity surveys follows.
\cite{1995MNRAS.276.1391N}  
presented a  method of deriving a  smoothed and unbiased estimated velocity  field from a survey of spiral galaxies with TF distances, based on  minimizing the scatter in the  inverse TF relation.
\cite{2015MNRAS.450.1868W} (hereafter WF15) 
address the bias by suggesting  a new estimator of the velocity, given the redshift and estimated distance. 
\cite{2015MNRAS.450.2644S} 
presented a different approach  to correcting the lognormal bias - a phenomenological correction to the 1-point distribution function of the velocities is introduced to eliminate the lognormal bias.   The procedure  was applied to the Cosmicflows-2 data set.
A fully Bayesian comprehensive approach  to the bias-free estimation of galaxy distances and velocities and to the (linear) reconstruction of the density and velocity  fields has been recently developed  \citep{2016MNRAS.457..172L,2019MNRAS.488.5438G}.  
The method is based on 
Markov Chain Monte Carlo (MCMC) sampling of  the posterior Bayesian probability distribution function and it has been applied to the CF3 data. A detailed comparison between the  method  presented here  and the  MCMC  approach of   \cite{2019MNRAS.488.5438G}
is to be presented elsewhere (Valade et al, in prep).

Application of the WF/CRs to the observed un-corrected CF3-like data so as to reconstruct the local underlying  density and  velocity fields would yield a very distorted structure. Consider, for example, the mock data presented  in Fig. \ref{fig:V-vs-D}. A WF reconstruction would give rise  to a spurious  radial outflow out to a distance of $\approx70\hmpc$ (where $h=H_0/100\kmsMpc$  and $H_0$ is Hubble's constant) and a strong inflow  beyond that. These false outflow and inflow are induced  by  the lognormal bias of the velocity data coupled with the given radial distribution of the data points. A close examination of how the WF works reveals that it is insensitive to sampling issues that give rise to the various aspects of Malmquist bias but is strongly affected by the lognormal bias. The aim of the present paper is to present a new scheme for correcting the lognormal bias within the framework of the WF/CRs reconstruction. This is tested here against CF3-like mock  catalogs drawn  from a cosmological  simulation constructed within the framework of the \LCDM\  standard  model of cosmology. 

The paper starts with a brief review of the CF3 data (\S \ref{sec:CF3}) and the constructed CF3-like mock catalogs (\S \ref{sec:CF3-like-mocks}). 
The methodology is  presented in \S \ref{sec:method},
and its testing against mock catalogs in \S \ref{sec:application}.
It is further tested in the context of the WF reconstruction from velocity surveys (\S \ref{sec:WF}) and of the estimation of Hubble's constant
(\S \ref{sec:H0-mocks}).
The bias corrected actual Cosmicflows-3 catalog is presented in \ref{sec:CF3-BGc}.
A summary and a discussion conclude the paper (\S \ref{sec:disc}).

\section{The grouped Cosmicflows-3 data}
\label{sec:CF3}

\cite{2016AJ....152...50T} provides a detailed account of the CF3 data. Given our interest in the linear WF/CRs reconstruction we focus  here on the grouped version of the CF3 data. The grouping process collapses all galaxies    that are  members of a given group or a cluster into a single data  point. This  suppresses the virial motions within collapsed halos and thereby effectively eliminates the main source of non-linearities in the velocity field. The error  of a grouped data point  is the mean of the errors on the distance moduli  of the members of the group  reduced by the inverse of the square root of the number of the group's  members. It follows that a data point of the grouped CF3 data is either identical to that of  the full CF3 data for  a group of a single member or it inherits the properties of its member galaxies by taking the arithmetics mean of their redshifts and angular positions with a reduced error. 

\begin{figure}
\centerline{
\includegraphics[width=1.0\columnwidth]{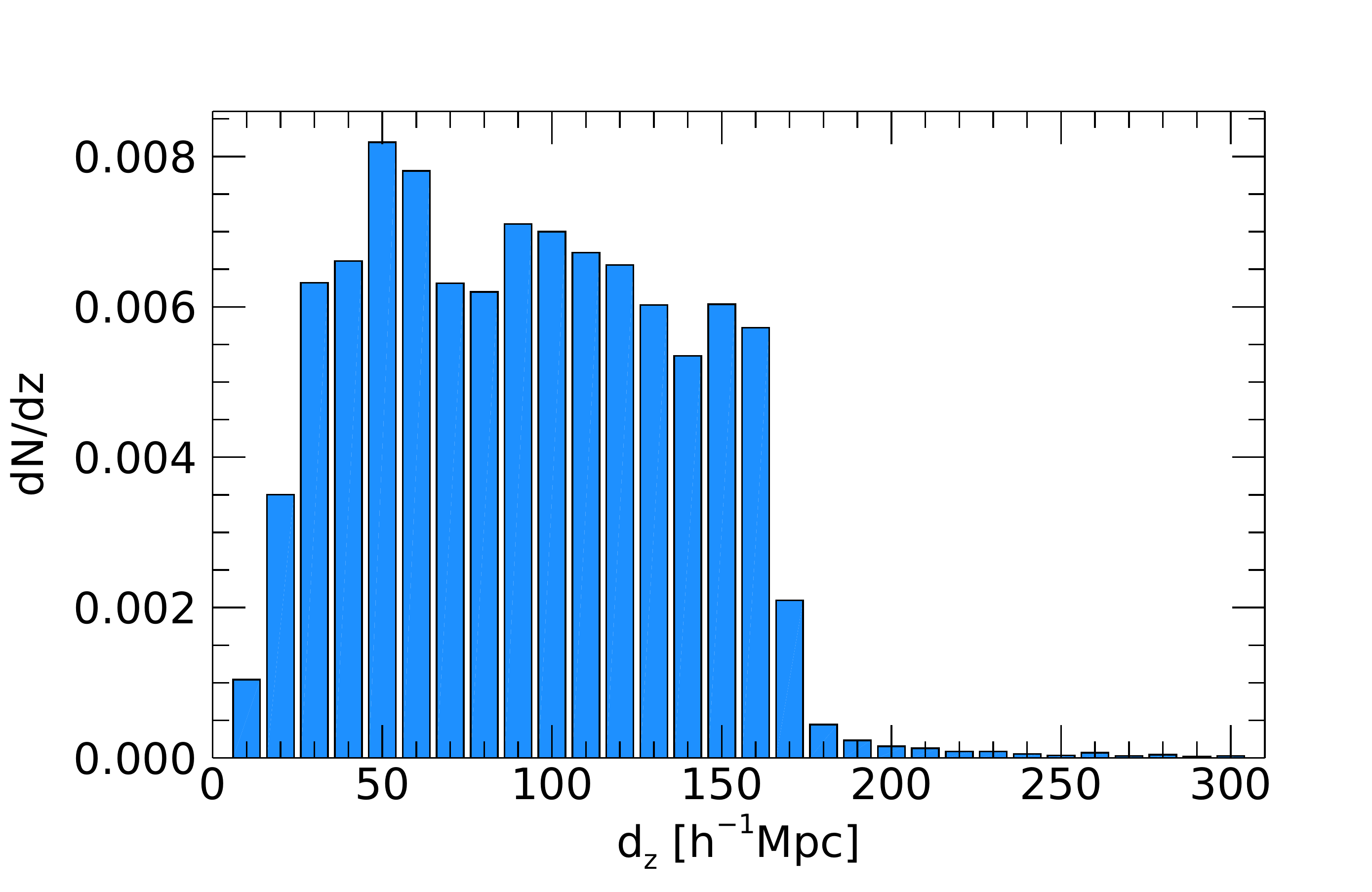} 
}
\caption{ CF3 grouped data: Density of the distribution of data point with respect to their  redshift distance ($d_z=cz/H_0$) 
}
\label{fig:CF3-histogram}
\end{figure}

The grouping  reduces the number  of entries from 17,669 down to 11,501. The process suppresses the  internal virial  motions within groups and clusters, and thereby  provides a proxy to their linear velocities.  The redshift  distance distribution of the data points is presented by Fig. \ref{fig:CF3-histogram}. The data manifests a sharp redshift cutoff at a redshift distance of $d_z \sim 160\ \hmpc$ ($d_z = c z / H_0$), with only $\sim$400 galaxies  that lie outside of that limit.

\section{Mock Cosmicflows-3 catalogs}
\label{sec:CF3-like-mocks}

Mock catalogs of the grouped CF3 data have been constructed so as to test the bias correction scheme  presented here.  The mock catalogs are drawn from  the publicly available Multi-Dark-2 simulation\footnote{see https://www.cosmosim.org} (Klypin et al 2016).  
    This is a dark matter only $N$-body simulation  with $N=3840^3$ in a periodic box of side length  $1.0 \hgpc$ particles of mass $m=1.5\times 10^{9}\hmsun$ with a force resolution of $\epsilon=5$\,kpc,  assuming a Planck cosmology ($H_0=67.7\kmsMpc$, $\Omega_{\Lambda}=0.69$, $\Omega_{\rm b}=0.04$, $\Omega_{\rm m}=0.31$ $\sigma_{8}=0.82$)  \citep{Planck:2013}.
 A ``Friends-of-Friends''  (FOF) algorithm is run on the $z=0$ particle distribution and all groups larger than 20 particles are retained as halos.
 By design all FOF halos are `parent' halos - namely ones that are not embedded in larger and more massive halos. Parent halos have substructures but are not  substructure themselves.

The data points of the grouped CF3 catalog correspond to collapsed groups and clusters of galaxies, some of which are made of single galaxies. The assumption made here is that these Cf3 groups - composed  of single or multiple galaxies -  are associated with parent halos of the DM particles distribution.  The construction of the mock grouped CF3 data  is based on the pairing of the actual observed data points with parent halos of the simulation, retaining their observed positions in redshift space and their errors. Thus, the mock data points are `ignorant' of the grouping of  the actual CF3 galaxies other than inheriting the errors of the grouped CF3 data points that are strongly affected by the grouping.
It follows that the issue of grouping is left out of the discussion  here and it is assumed to be done prior to the analysis presented in the paper.

The construction  starts with selecting a Milky Way mass halo as the mock LG, namely, one halo with mass $9\times10^{11}M_{\odot}< M_{\rm cen} <2\times10^{12}M_{\odot}$ (out of 3,333,049 possibilities) is randomly chosen to host the mock observer. Using the periodicity of the simulation, all haloes are centered on the mock observer.  A  Supergalactic-like cartesian coordinate system  aligned with the computational box and centered on the mock LG,  is then constructed with the (SGX, SGY, SGZ) orientation randomly selected.
 
The procedure aims at pairing simulated halo with CF3 data points, while minimizing their separation distances in redshift space.   This is done in two steps, per each CF3 data point. Halos are then projected on to the mock sky, and are binned in Supergalactic  (SGL, SGB) bins. All the halos in the angular bin of a given CF3 data point are identified, and among them the halo closest in redshift to the observed redshift is selected for the mock catalog. The estimated errors of the CF3 data are assigned to their mock counterparts.

The algorithm of generating mock catalogs includes the  application of the Friedmann-Robinson-Walker (FRW) metric to the otherwise Newtonian variables of the numerical code, turning Newtonian spatial coordinates into luminosity distances and  the simple Newtonian peculiar velocities into the ones actually observed in an FRW universe.

Ten different mock observers have been randomly selected  and ten different errors realizations have been constructed yielding 100 different mock catalogs altogether. The  ten different observers sample the cosmic variance of CF3-like surveys and the ten errors realizations sample the error variance associated with the observational uncertainties associated with the individual measurements of individual data points.
All the different scatter plots shown in the paper are based on data drawn from one of these mock catalogs. 

\section{Methodology}
\label{sec:method}

\subsection{Basics}

Peculiar velocity surveys/catalogs consist of  the luminosity distance, $d_L$,  the observed redshift, $z$, angular position and the errors associated with the luminosity distance  and the redshift of a collection of  data points. 
All the steps involved in extracting proper distances ($d$) and peculiar radial  velocities ($v$) from Cosmicflows-like data are briefly presented here. 
The following derivation  is given here for the sake of completeness  
\citep[see][]{2008cosm.book.....W,2014MNRAS.442.1117D}.

In a flat  universe the cosmological redshift $z_{\rm cos}$ of the luminosity distance  are related by:
\begin{equation}
d_L=c H{^{-1}_0}(1+\zcos)  \int_0^{\zcos}{1\over\sqrt{\Omega_m(1+z)^3+(1-\Omega_m)}}\d{z}
\label{eq:dL-zcos}
\end{equation}
(here $c $ and $\Omega_m$ are the speed of light, and the cosmological matter density parameter, respectively). Eq. \ref{eq:dL-zcos}  is solved to find $z_{\rm cos}$ for a given $d_L$ and a given cosmology ($H_0$ and $\Omega_m$). The proper  distances ($d$)  is given by:
\begin{equation}
d={d_L\over (1 + \zcos) }
\label{eq:d-dL}
\end{equation}

The observed redshift of an object is related to its cosmological redshift and peculiar radial velocity ($v$)  by,
\begin{equation}
 1+z = (1+\zp)(1+\zcos),
\label{z-zcos-zp}
\end{equation}
where $z_{\rm p}=v/c$. 
Eq. \ref{z-zcos-zp} is used to calculate $v$. A 1st order expansion of Eq. \ref{z-zcos-zp} yields the familiar expression,
\begin{equation}
cz\sim H_0 d +v.
\label{cz-H0d-v}
\end{equation}

For the vast majority of the galaxies in the CF3 catalog - e.g. spirals with Tully-Fisher distances  and ellipticals with FP distances -  distance  estimation from observables starts with the   distance moduli ($\mu$).  It is related to the (luminosity) distance by
\begin{equation}
\mu = 5 \log_{10}\left({d_ L\over 10 {\,\rm pc}}\right)
\label{eq:mu-d}
\end{equation}
For mathematical convenience Eq. \ref{eq:mu-d} is rewritten as 
\begin{equation}
\mu  = B \ln\left({d_L\over A}\right)
\label{eq:mu-d-ln}
\end{equation}
where $B=5/\ln{10}$ and $A=10 \,{\rm pc}$.

It is commonly assumed  that errors on the observed distance modulus are normally distributed, $\mu_{obs}=\mu +\sigma_\mu \epsilon$, where $\epsilon\in\mathcal{N}(0,1)$ (namely, normal distribution of zero mean and variance of unity). It follows that the `observed' luminosity distance is
\begin{equation}
D_L=A\exp({\mu_{\rm obs}\over B})=A\exp({\mu +\sigma_\mu \epsilon\over B})
\label{eq:dobs}
\end{equation}
A first order expansion   of the  observed with respect to the true proper distances, namely $D$ with respect to $d$, yields:
\begin{equation}
D \approx d (1 + \tilde{\sigma}_\mu \epsilon)
\label{eq:d-1st-order}
\end{equation}
where 
$\tilde{\sigma}_\mu=\sigma_\mu/B$.
Here $d$ and $D$ are the true and observed proper distances of a data point, respectively.
It is noted here that the distance measurement errors scale with true    and not with the observed distance.
The typical uncertainty in the TF relation for a field spiral is $\sigma_\mu\sim 0.4$ resulting in fractional (TF) distance uncertainty of $\sim 0.18$.
The observational uncertainty on the derived velocity, $\epsilon_v$ is closely related to the distance uncertainty,
\begin{equation}
\epsilon_v=-H_0 d \tilde{\sigma}_\mu \epsilon.
\label{eq:eps-v}
\end{equation}
Here  the $50\kms$ uncertainties on the redshifts are neglected.

The normal distribution of  the distance modulus errors  and the exponential dependence of the distance on the distance modulus imply that the distance and  the velocity errors are lognormal  distributed. Hence the distance and velocity estimations are biased. Namely,  even in the limit of infinite many repeated observations of a give galaxy the means of the estimated distances and velocities do  not coincide with their actual values.

\begin{figure}
\centerline{
\includegraphics[width=1.0\columnwidth]{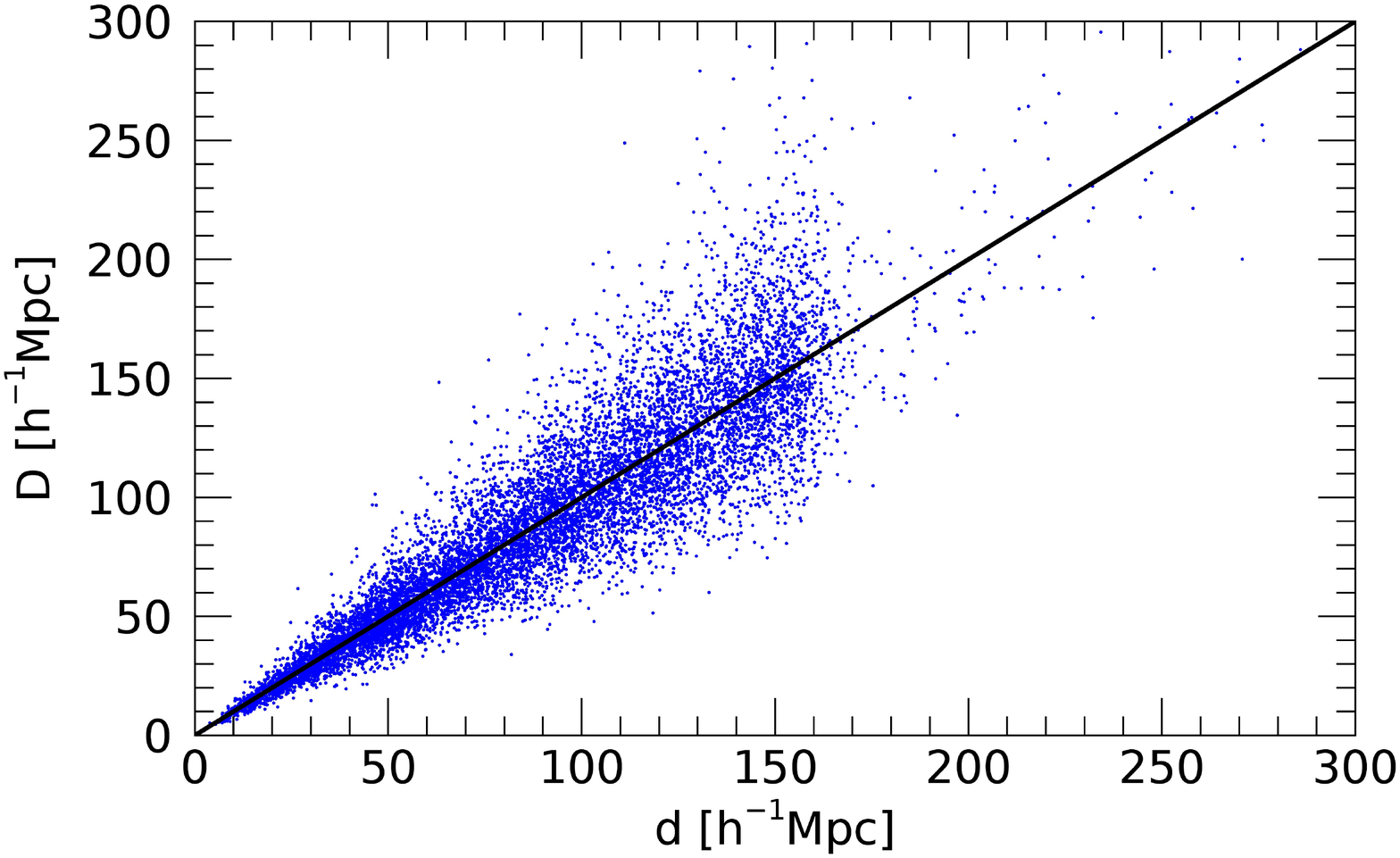} 
}
\caption{ A scatter plot of the observed ($D$) vs the actual ($d$) distances drawn from a CF3-like mock catalog. 
}
\label{fig:dobs-vs-d}
\end{figure}

\begin{figure}
\centerline{
\includegraphics[width=1.\columnwidth]{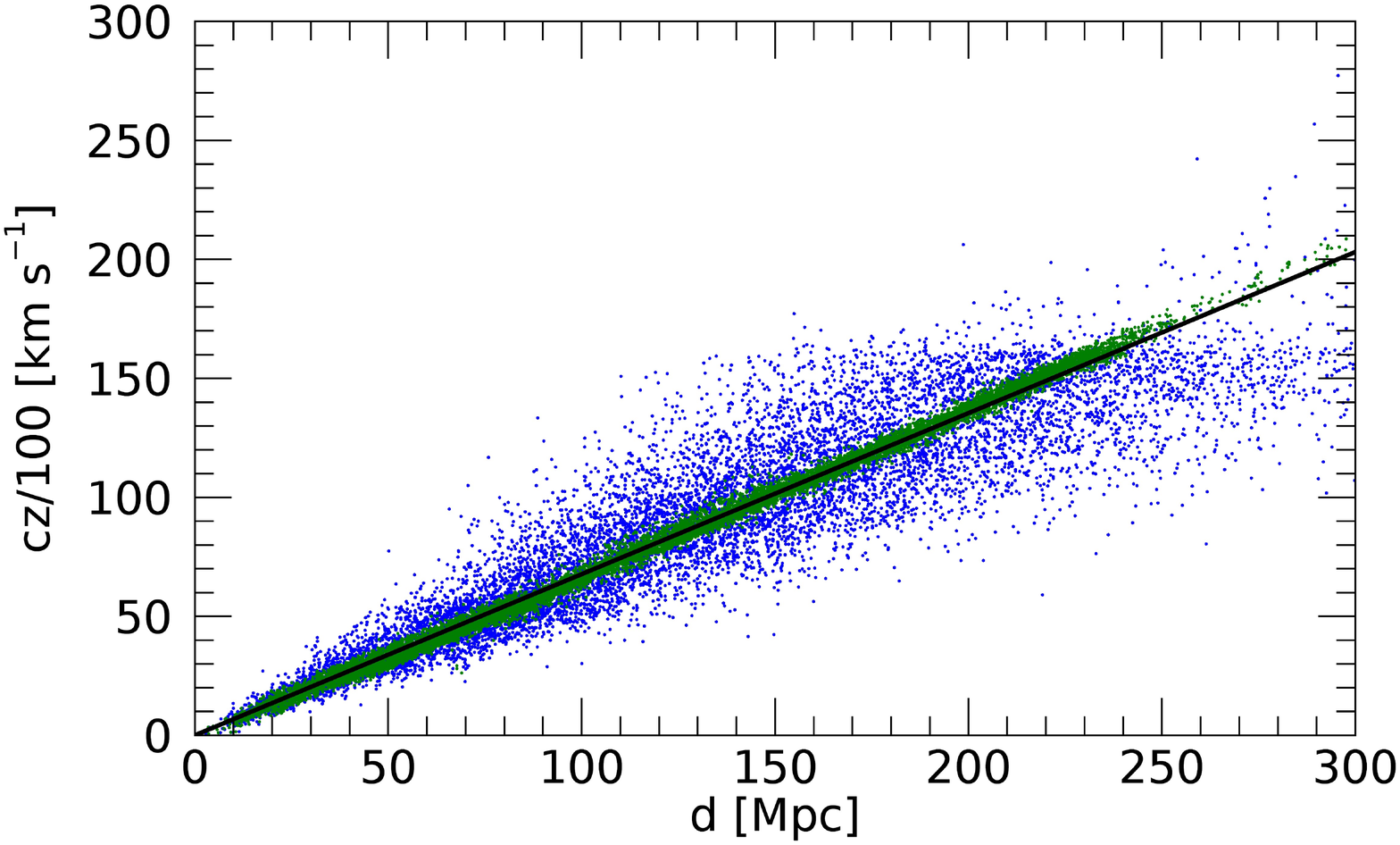} 
}
\caption{ A Hubble diagram, namely $cz$ vs. distance, is drawn for one of the mock catalogs.  Distances are given in Mpc, taking the actual $H{^{\rm \Lambda CDM}_0}=67.7 \kmsMpc$ of the simulation. Blue symbols correspond to observed data points and green ones for the actual,   values. The solid line corresponds to the Hubble velocity-distance law.
}
\label{fig:Hubble-diag}
\end{figure}

 The exponential dependence of the distance on the distance modulus makes the distribution of the estimated errors  skewed (Fig. \ref{fig:dobs-vs-d}). That bias affects the distribution of the estimated peculiar velocities with respect to  their estimated distances in two ways - it changes the spatial distribution of the data points and the actual values of the estimated velocities.
This is clearly depicted  by Fig. \ref{fig:V-vs-D} which shows the distribution of velocities and distances of the observed   mock data points. The bias  affects also a Hubble diagram constructed from the mock catalog, namely $c z$ vs the distances of the mock data. Eq. \ref{cz-H0d-v} implies that a small scatter is expected even for the 'clean' (i.e zero errors) data, as a consequence of the emergence of peculiar velocities associated with structure formation out of a primordial density perturbation field (Fig. \ref{fig:Hubble-diag}). For the standard \LCDM\ model assumed here the expected scatter is roughly $\sigma_{vp}\sim 275 \kms$ (for the radial component of the velocity as is calculated from the simulated mock catalogs).

In the standard model of cosmology peculiar velocities are assumed to constitute a random (vector) Gaussian field defined by the primordial power spectrum. Non-linear dynamics and in particular virial motions in collapsed groups and clusters of galaxies violate the prediction of the linear theory of structure formation. Grouping the CF3 catalog  suppresses the non-linear motions and acts as an effective linearization of the observed velocities. It follows that  radial velocities are expected to scatter normally as a function of their distance from a random observer in a \LCDM\ universe. Fig. \ref{fig:V-vs-D} shows the scatter of    actual   $v$ vs $d$ and of   observed  $V$ vs $D$ for one of the mock catalogs. The bias is very clearly manifested.

\subsection{Distances}
\label{subsec:BGc-d}
We first consider an ideal case of a subsample of data points all of which have  the same actual distance ($d$) but   different observed distances ($D$). 
The lognormal  probability distribution function (PDF)  of the observed given the actual distances is given by
\begin{equation}
P(D\vert d)={1\over \sqrt{2\pi}\tilde{\sigma}_\mu} \exp\Big( - {\big(\ln(D/d)\big)^2\over 2 \tilde{\sigma}{_\mu^2}}   \Big) {1\over D}.
\label{eq:D-d}
\end{equation}

Given the lognormal PDF  (Eq. \ref{eq:D-d}) one finds that the median, mean, variance and standard deviation of $D$   given $d$ are easily calculated: 
\begin{eqnarray}
(D|d)_{med} & = &  d  \nonumber\\
< D|d > & = &  d \exp({\tilde{\sigma}{^2_\mu}\over 2} ) \nonumber\\
< D^2|d > & = &  d^2 \exp(2\tilde{\sigma}{^2_\mu} )                     \\
\sigma_D & = & d \sqrt{\exp(2\tilde{\sigma}{^2_\mu} )  - \exp(\tilde{\sigma}{^2_\mu}) } \nonumber
\label{eq:mean-D}
\end{eqnarray}
Here $\sigma{^2_D}$ is the central moment of  $D$ given $d$.

The  value of  $d$  can be estimated from the  subsample by  recalling  that median of $D$ equals the actual distance 
(Eq. 11). 
It follows that $(D|d)_{med} $ is the distance estimator for all observed distances $D$. Eq. \ref{eq:D-d} is now rewritten as:
\begin{equation}
P(D\vert d)={1\over \sqrt{2\pi}\tilde{\sigma}_\mu} \exp\big( -{(\ln{D/(D|d)_{med}})^2\over 2 \tilde{\sigma}{_\mu^2}}   \big) {1\over D},
\label{eq:D-Dmed}
\end{equation}
Expressing $D$ in terms of $\epsilon$  and recalling Eqs. \ref{eq:mu-d-ln} and  \ref{eq:dobs} one  writes:
\begin{equation}
\epsilon(D)={1\over\tilde{\sigma}_\mu}\ln({D\over  (D|d)_{med}})
\label{eq:epsilon-D}
\end{equation}

The true distance, $d$, is unknown, but the redshift distance, defined here as $d_z=cz_{obs}/H_0$,  is very accurately determined, and therefore we seek to replace the former by the latter:
\begin{eqnarray}
P(d,d_z) &=& P(d_z|d) P(d)                     \\
              &=& P(d|d_z) P(d_z),   \nonumber
\label{eq:PdPdz-1}              
\end{eqnarray}
where $P(d)$ is the probability of finding a data point in in an interval 
$\d{d}$,   $P(d)\d{d} = n(d)$  , where $n(d)$ is the number of data points in that interval. $P(d_z)$ is similarly defined, $P(d_z)\d{d_z}=n(d_z)$. 
The derivations and discussion  that follow strictly apply to the case of the grouped CF3 data where $P(d_z)\approx const.$ over the range of $30 \lesssim d_z \lesssim 160 \hmpc$ (see
Fig. \ref{fig:CF3-histogram}). 
This should also be  the case for $P(d)$ given that over that range 
$(d - d_z)_{rms} \approx 2.75 \hmpc $. Indeed,  the analysis of the mock CF3 data that follows  corroborates  this assumption. 

It follows that  
\begin{equation}
P(d|d_z) \approx  P(d_z|d).
\label{eq:PdPdz-2}              
\end{equation}
From here on this approximate relation is taken as an equality.

Next, Eq. \ref{eq:D-Dmed} is rewritten here as the conditional probability of $D$ given the redshift distance: 
\begin{eqnarray}
P(D|d_z)& &=\int P(D|\tilde{d}) P(\tilde{d}|d_z) \d{\tilde{d}} =  \\
               & &{1\over \sqrt{2\pi}\tilde{\sigma}_\mu D} {H_0\over \sqrt{2\pi}\sigma_{vp}} 
                \int\exp\Big( -{\big(\ln(D/(D|\tilde{d})_{med})\big)^2\over 2 \tilde{\sigma}{_\mu^2}}   \Big)  \nonumber \\
               & &\exp\big( -{(\tilde{d}-(D|z)_{\rm med})^2\over 2 (\sigma_{vp}/H_0)^2}   \big) \d{\tilde{d}}.  \nonumber
\label{eq:PdPdz-3}               
\end{eqnarray}
(Here $\tilde{d}$ is an integration variable.)

In the limit where  $\sigma_{vp} \ll cz$ the following holds,
\begin{equation}
P(D\vert d_z)={1\over \sqrt{2\pi}\tilde{\sigma}_\mu} \exp\big( -{(\ln{D/(D|z)_{\rm med}})^2\over 2 \tilde{\sigma}{_\mu^2}}   \big) {1\over D},
\label{eq:D-z}
\end{equation}
The following analytical derivations are made under that assumption. In particular, Eq. \ref{eq:epsilon-D} is rewritten here as
\begin{equation}
\epsilon(D)={1\over\tilde{\sigma}_\mu}\ln({D\over  (D\vert z)_{\rm med}}).
\label{eq:epsilon-D-z}
\end{equation}

Our aim here is to map the lognormal distributed $D$ to an estimator, $d_{BGc}$ where BGc stands for bias gaussianization correction,  that is normally distributed  around the actual distance, $d$,  with an unspecified yet  dispersion of $\sigma_d$. Namely, 
\begin{equation}
P(d_{BGc}\vert d_z)={1\over \sqrt{2\pi}\sigma_d} \exp\big( -{(d_{BGc}-(D\vert z)_{\rm med})^2\over 2 \sigma{_d^2}}   \big).
\label{eq:dG-z}
\end{equation}

Given the lognormal transformation from the true to the observed distance, the cumulative distribution of $D$ (Eq. \ref{eq:D-d})  equals to that of $d$ (Eq. \ref{eq:dG-z}). It follows that the following approximate equality determines $d_{BGc}$:\\
\begin{equation}
{1\over\sqrt{2\pi}}\int_{-\infty}^{\epsilon(D)}  \exp(-{\epsilon^2\over 2}) \d{\epsilon} =
\int_0^{d_{BGc}} P\big( \tilde{d}_G\vert d_z)\d{\tilde{d}_G},
\label{eq:dg-epsilon}
\end{equation}
where the PDF in the integrand of the RHS of the equation is given by Eq.   \ref{eq:dG-z}. 
Over the vast majority of the CF3 data points the RHS of Eq. \ref{eq:dg-epsilon} can be very well approximated by:
\begin{multline}
{1\over\sqrt{2\pi}}\int_{-\infty}^{\epsilon(D)}  \exp(-{\epsilon^2\over 2}) \d{\epsilon} =\\
{1\over \sqrt{2\pi}\sigma_d}  \int_{-\infty}^{d_{BGc}} 
                 \exp\big( -{(\tilde{d}_G-(D\vert z)_{\rm med})^2\over 2 \sigma_{_d}^2}   \big)\d{\tilde{d}_G},
\label{eq:dg-epsilon-infty}
\end{multline}
The solution of Eq. \ref{eq:dg-epsilon-infty} is  trivial,
\begin{equation}
{\rm erf}({1\over\sqrt{2}\tilde{\sigma}_\mu}\ln({D\over  (D\vert z)_{\rm med}}))={\rm erf}({d_{BGc}-(D\vert z)_{\rm med}\over\sqrt{2} \sigma_d})
\label{eq:erf-dg}
\end{equation}
where ${\rm erf}(x)$ is  the error function. It follows that
\begin{equation}
d_{BGc}=(D\vert z)_{\rm med} + {\sigma_d\over\tilde{\sigma}_\mu}\ln{\big({D\over(D\vert z)_{\rm med}}}\big).
\label{eq:dg}
\end{equation}

 Eq. \ref{eq:PdPdz-3}  is derived under the assumption that the PDF of the peculiar velocities of the data points is Gaussian. This is the standard assumption of the \LCDM\ model as far as the velocity field of the DM particles is concerned. Yet, the sampling of the velocity field by collapsed groups of the CF3 galaxies  leads to non-linear coupling between the density and velocity fields, and this might lead to deviations from a Gaussian distribution. The construction of the mock data by associating the grouped CF3 data with parent halos of the DM particles distribution properly addresses this concern. An inspection of the PDF of the velocities of the mock data points shows that it is very well approximated by a Gaussian.

\subsection{  Peculiar  Velocities}
\label{subsec:BGc-v}
 
 Again, we start here with  a  subsample of data points of  a given true distance, $d$. Eq. \ref{cz-H0d-v} is assumed here and  the   observed  peculiar  velocity ($V$) is
\begin{equation}
V =  cz -H_0 D
\label{V-cz-H0-D}
\end{equation}
(to 1st order in $v/cz$).

Given the lognormal PDF  (Eq. \ref{eq:D-d}) and similarly to Eq. \ref{eq:mean-D} one finds that the median, mean, variance and standard deviation of $V$   given $d$ are easily calculated: 
\begin{eqnarray}
(V|d)_{med} & = & cz - H_0 d  \nonumber\\
< V|d > & = &  cz - H_0 d \exp({\tilde{\sigma}{^2_\mu}\over 2} ) \\
\sigma_V & = & H_0d \sqrt{\exp(2\tilde{\sigma}{^2_\mu} )  - \exp(\tilde{\sigma}{^2_\mu}) }  \nonumber
\label{eq:mean-V}
\end{eqnarray}
Here $\sigma{^2_V}$ is the central moment of  $V$ given $d$.
It depends on the unknown true distance but  it can be approximated by
\begin{equation}
\sigma_V \sim cz\sqrt{\exp(2\tilde{\sigma}{^2_\mu} )  - \exp(\tilde{\sigma}{^2_\mu}) }.
\label{eq:mean-cz-V}
\end{equation}
A 1st order expansion yields
\begin{equation}
\sigma_V \sim  cz \tilde{\sigma}_\mu.
\label{eq:sigma-cz-V}
\end{equation}

The median of the observed velocities, $(V|d)_{\rm med}$, and of the distances of the  subsample are related by:
\begin{equation} 
(V|d)_{\rm med}=cz-H_0 (D|d)_{med}=cz-H_0 d
\label{eq:Vmed-v}
 \end{equation}
It follows that  $(V|d)_{\rm med} =v$ (the actual peculiar  velocity).
The errors on the distance lead to errors on the velocities by
\begin{equation}
V=(V|d)_{\rm med}-H_0 d\big(\exp(\tilde{\sigma}_\mu)-1\big).
\label{eq:Vmed-d-err}
\end{equation}

Replacing the ensemble of data points to one of a given redshift and under the assumption that the distance errors (scaled by $H_0$) are much larger than $\sigma_{vp}$, $H_0 d$ is replaced by $cz$ in Eq. \ref{eq:Vmed-d-err}. The normal error variable ($\epsilon$) is related to  the observed velocity by:\\
\begin{eqnarray}
\epsilon(V)&=&{1\over\tilde{\sigma}_\mu}\ln({cz -V\over cz - (V\vert z)_{\rm med}})\\
                 &=&{1\over\tilde{\sigma}_\mu}\ln({D\over  (D|d)_{med}}), \nonumber
\label{eq:epsilon-V}
\end{eqnarray}
where $V=cz-H_0D$.

Given the lognormal nature of the errors on the observed velocities we introduce a Gaussian estimator of the velocity, $v_{BGc}$, which scatters normally around the median with a dispersion that equals the errors of the observed velocities. Namely, the PDF of $v_{BGc}$ is
\begin{equation}
P(v_{BGc}\vert z)={1\over\sqrt{2\pi}\sigma_V}\exp\big(-{(v_{BGc}-V_{\rm med})^2\over 2 \sigma{_V^2}}\big)
\label{eq:P-vG}
\end{equation}
The velocity equivalent to   Eq. \ref{eq:dg-epsilon} is\\
\begin{equation}
{1\over\sqrt{2\pi}}\int_{-\infty}^{\epsilon(V)}  \exp(-{\epsilon^2\over 2}) \d{\epsilon}  =
\int_{-\infty}^{v_{BGc}} P\big( \tilde{v}_G\vert z)\d{\tilde{v}_G},
\label{eq:vG-V}
\end{equation}
whose  solution is again trivially given by\\
\begin{multline}
2-{\rm erfc}\big({1\over\tilde{\sigma}_\mu}\ln( {cz -V\over cz - (V\vert z)_{\rm med}} ) \big)=\\
{\rm erfc}\big({v_{BGc}-(V\vert z)_{\rm med}\over 2\sigma_V}  \big), 
\label{eq:erfc-vG}
\end{multline}
or
\begin{equation}
v_{BGc}=(V\vert z)_{\rm med} + 2\ \sigma_V \ {\rm erfc}^{-1}
\Big(2-{\rm erfc}\big(  {\ln({cz -V\over cz - (V\vert z)_{\rm med}} ) \over 2\tilde{\sigma}_\mu}\big) \Big)
 \label{eq:vg-1}
 \end{equation}
 Here ${\rm erfc}(x)$ is the complimentary error function and ${\rm erfc}^{-1}(x)$ is its inverse. 
 
 Given that Eq. \ref{cz-H0d-v} holds also for the medians of the distribution of $D$ and $V$, at a given redshift, namely that 
  \begin{equation}
cz =  H_0 (D\vert z)_{\rm med} +(V\vert z)_{\rm med}, 
\label{cz-H0d-v-med}
\end{equation}
and recalling that $ {\rm erfc}^{-1}\big(2-{\rm erfc}(x)\big)=-x$, 
Eq. \ref{eq:vg-1} is rewritten as:
\begin{equation}
v_{BGc}=(V\vert z)_{\rm med} - {\sigma_V  \over  \tilde{\sigma}_\mu}  
  \ln({D\over  (D\vert z)_{\rm med}} ) 
 \label{eq:vg}
\end{equation}

A comforting relation that emerges from Eqs. \ref{eq:dg}   and \ref{eq:vg} is the redshift, distance and velocity relation is valid also for the mean of the BGc estimated distances and velocities at a given redshift:
\begin{equation}
H_0 \large<  (d_{BGc}\vert z) \large>  + \large< (v_{BGc} \vert z) \large> =  cz.
\label{eq:dG-vG-cz}
\end{equation}
Namely, Eq. \ref{eq:dG-vG-cz} is the Gaussian equivalent of the non-Gaussian Eq. \ref{eq:Vmed-v}.

\subsection{Bias Gaussianization correction: practical implementation}
 \label{subsec:BGc}

A key  element in the  expressions for $d_{BGc}$ and $v_{BGc}$ is the evaluation of the medians of the observed distances and velocities for a given  redshift ($(D\vert z)_{med}$ and $(V\vert z)_{med}$). This depends on a construction  of a subsample of data points with redshifts close enough to  $z$.    The following reasoning is used to construct such a subsample. Define  the redshift  range  of the data points   where: a.  the observed radial velocities and the residual of redshift distances from the true ones are dominated by the distance measurements errors; b.   the sampling in redshift  is dense enough. A given data point within that range is considered here.   A subsample of $N_z$  data points adjacent to it in their redshifts - the closest in redshits $N_z/2$ data points smaller than $z$, and likewise for the ones larger than $z$. The medians of the distances and velocities are constructed for that particular data point. Eqs. \ref{eq:dg} and \ref{eq:vg} are  used to evaluate  the Gaussianized estimators $d_{BGc}$ and $v_{BGc}$, based on the medians evaluated for that particular data point. The procedure is repeated to all the data point  within that  redshift range. The key assumption made here, upon which the BGc algorithm is built, is that for the data points within that range and their associated subsample  the residual between observed and true distances and velocities are dominated by  uncertainties, i.e. `errors', on the distance moduli and that these are uncorrelated. It is this statistical independence of the errors on $\mu$ which enables one to construct the subsample by their proximity in redshift regardless of their distances in the 3D redshift space.

For the case of the CF3 data and the \LCDM\  model this range corresponds to $15\  \lesssim\  d_z\ \lesssim\ 160\  \hmpc$. The lower bound is derived under the assumption that  $\sigma_{vp}\sim 275 \kms$ and that for the nearby data points $\large<\tilde{\sigma}_\mu\large> \sim  0.19$, or $\large<d_z H_0 \tilde{\sigma}_\mu\large> \sim  285 \kms$. As for the upper bound, the sampling below $ d_z\ \lesssim\ 160\  \hmpc$ is roughly uniform and is relatively extremely sparse beyond it (Fig. \ref{fig:CF3-histogram}). For the data points outside that range the following is applied: a. for the nearby data points the observed distances and velocities are retained; For the few distant data points observed distances are replaced by their redshift distances and their velocities are retained.

\begin{figure}
\centerline{
\includegraphics[width=1.0\columnwidth]{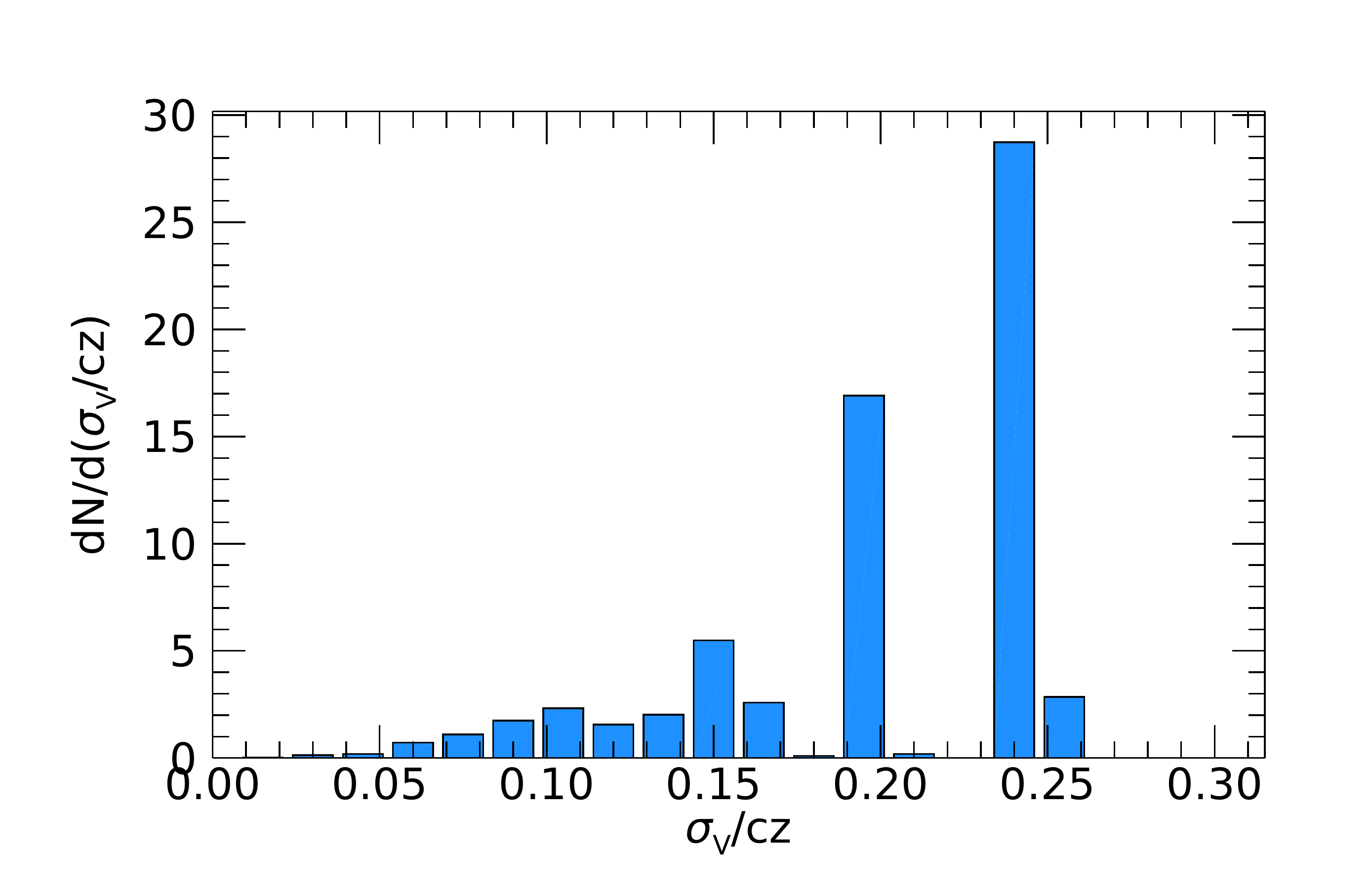} 
}
\caption{
CF3 data points: Density of the distribution of data point with respect to their   fractional distance errors.
}
\label{fig:histogram-err} 
\end{figure}

To the extent that the data is composed of galaxies or groups of  galaxies of  a given redshift and   all have the  same fractional errors then Eqs. \ref{eq:dg} and \ref{eq:vg} provide  accurate BGc estimations of the distances and velocities of the data points.  However, this is not the case with the actual CF3 survey, where the   two major subsets of data have fractional distance errors of $\tilde{\sigma}_\mu\sim0.18$ (TF galaxies) and $\sim0.24$ (6dF FP galaxies) (Fig. \ref{fig:histogram-err}). The grouping of the data adds an additional scatter to the distance errors. It follows that Eqs. \ref{eq:dg}  and  \ref{eq:vg} are only approximately correct - an approximation that is used in the rest of the paper. We have found here that the best - in the sense discussed below - BGc estimation is obtained here by replacing the actual values of $\tilde{\sigma}_\mu$ of individual data points by their mean value of $\tilde{\sigma}_\mu=0.19$.

The BGC estimated distances are obtained by the adding  a normally distributed term  to the median $(D|z)_{med}$ normalized by $\sigma_d/\tilde{\sigma}_\mu$ (Eq. \ref{eq:dg}). The normally distributed term keeps the median and  the mean of the distribution of $d_{BGc}$ invariant and roughly equal. We take here $\sigma_d=0$ and thereby  minimize the  scatter of the BGc around the redshift distances to its minimal value of $\sim\sigma_{vp}/H_0$.

\section{Application to CF3 mock catalogs}
\label{sec:application}
 
 The BGc procedure is examined here for one of the mock catalogs. The results are representative of all the other mock data sets.

\begin{figure}
\centerline{
\includegraphics[width=1.0\columnwidth]{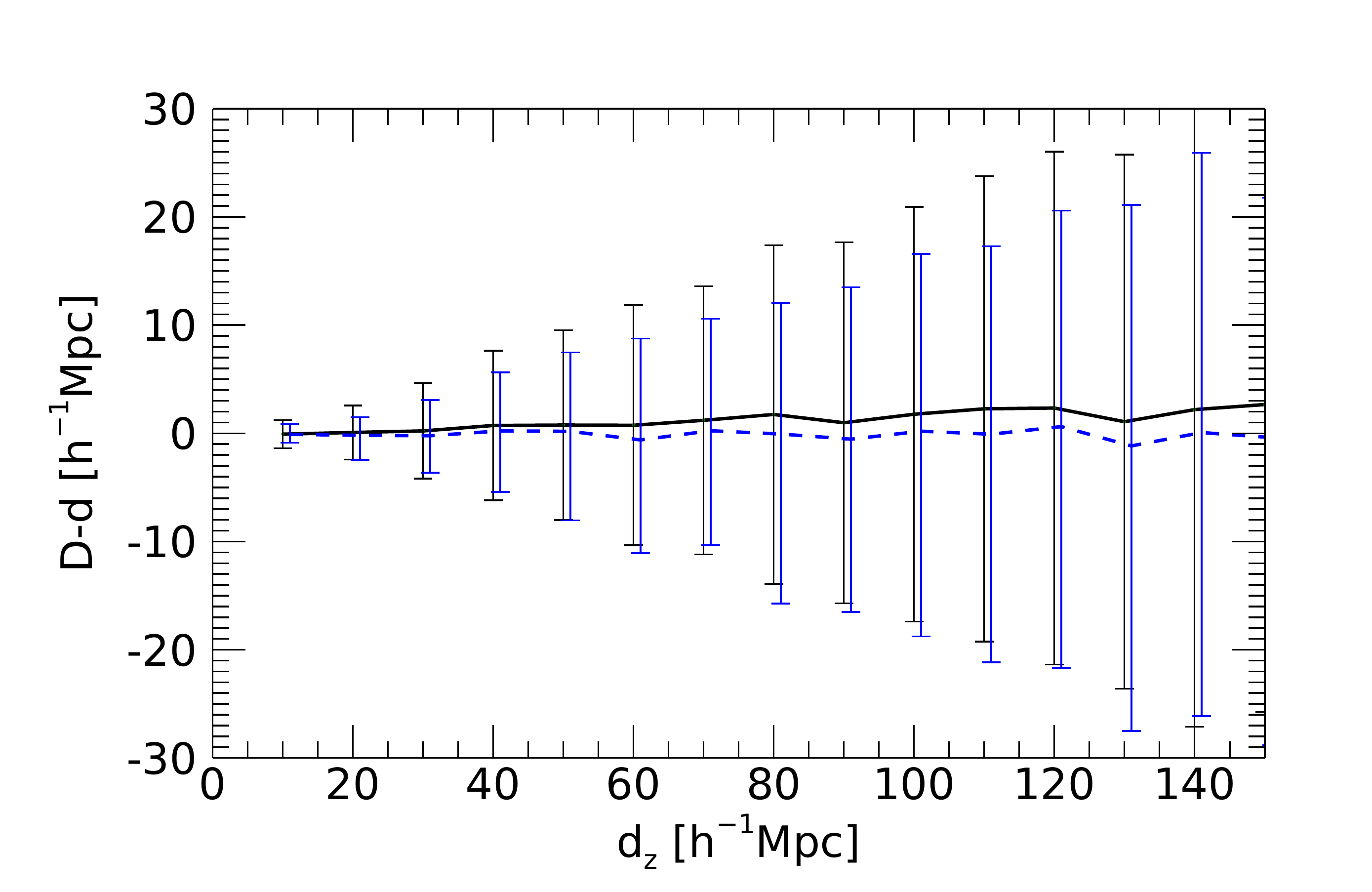}
}
\centerline{
\includegraphics[width=1.0\columnwidth]{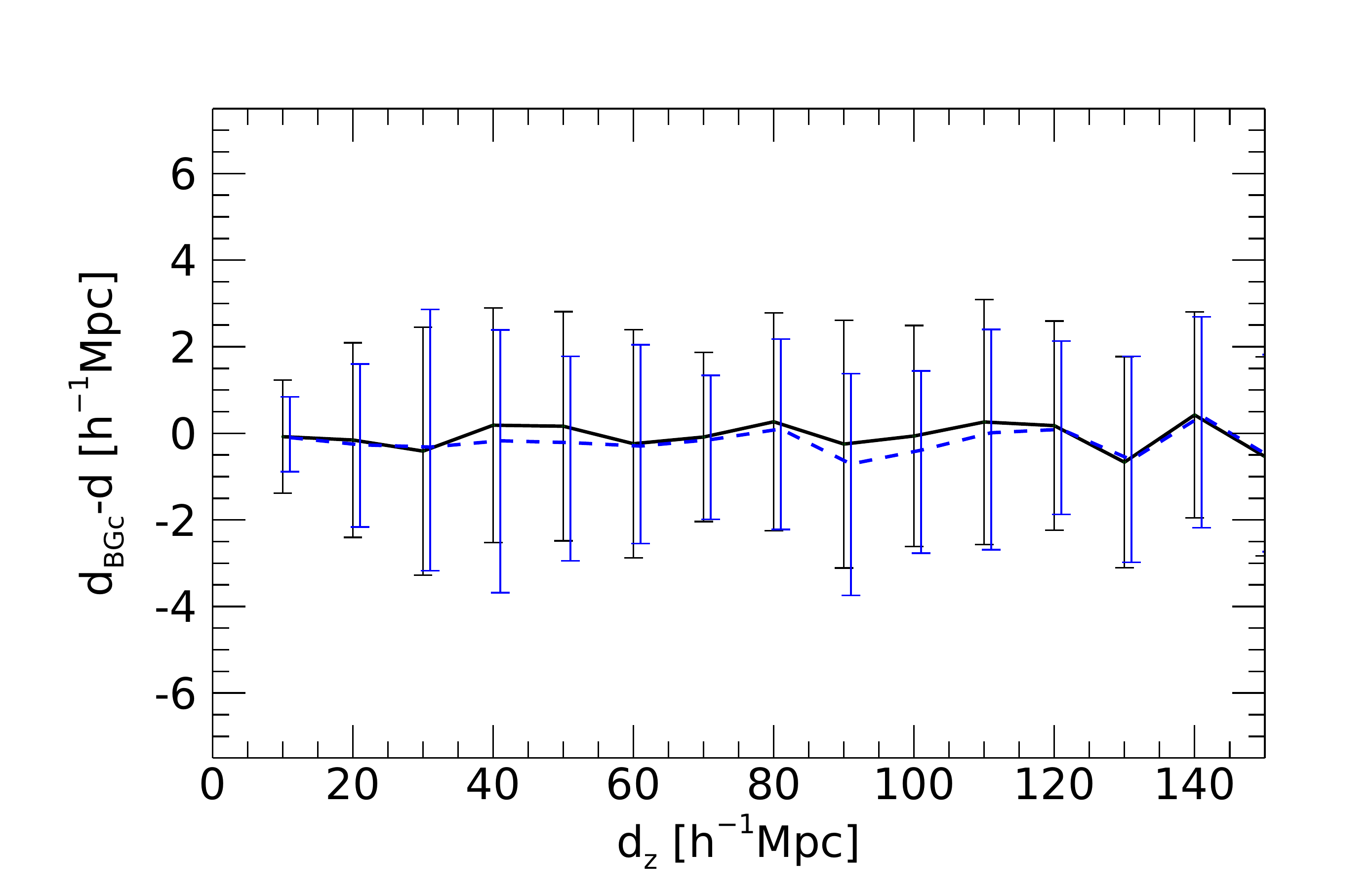}
}
\caption{ 
Mean (black, solid line) and median (blue, dashed line) 
of the residual of the observed distances (D) (top panel) 
and of the  estimated distances ($d_{BGc}$) (lower panel) from the actual distances ($d$) vs the redshift distances ($d_z$) for the mock data points. 
Note the different vertical scales of the  plots. (Black error bars correspond to one standard deviation around the mean value and the blue ones to the 1st and 3rd quartiles around the median, rescaled to correspond to the standard deviation of a normal distribution. The blue error bars are slightly shifted horizontally.) 
} 
\label{fig:dBGc-D-residual-dz} 
\end{figure}

\begin{figure}
\centerline{
\includegraphics[width=1.0\columnwidth]{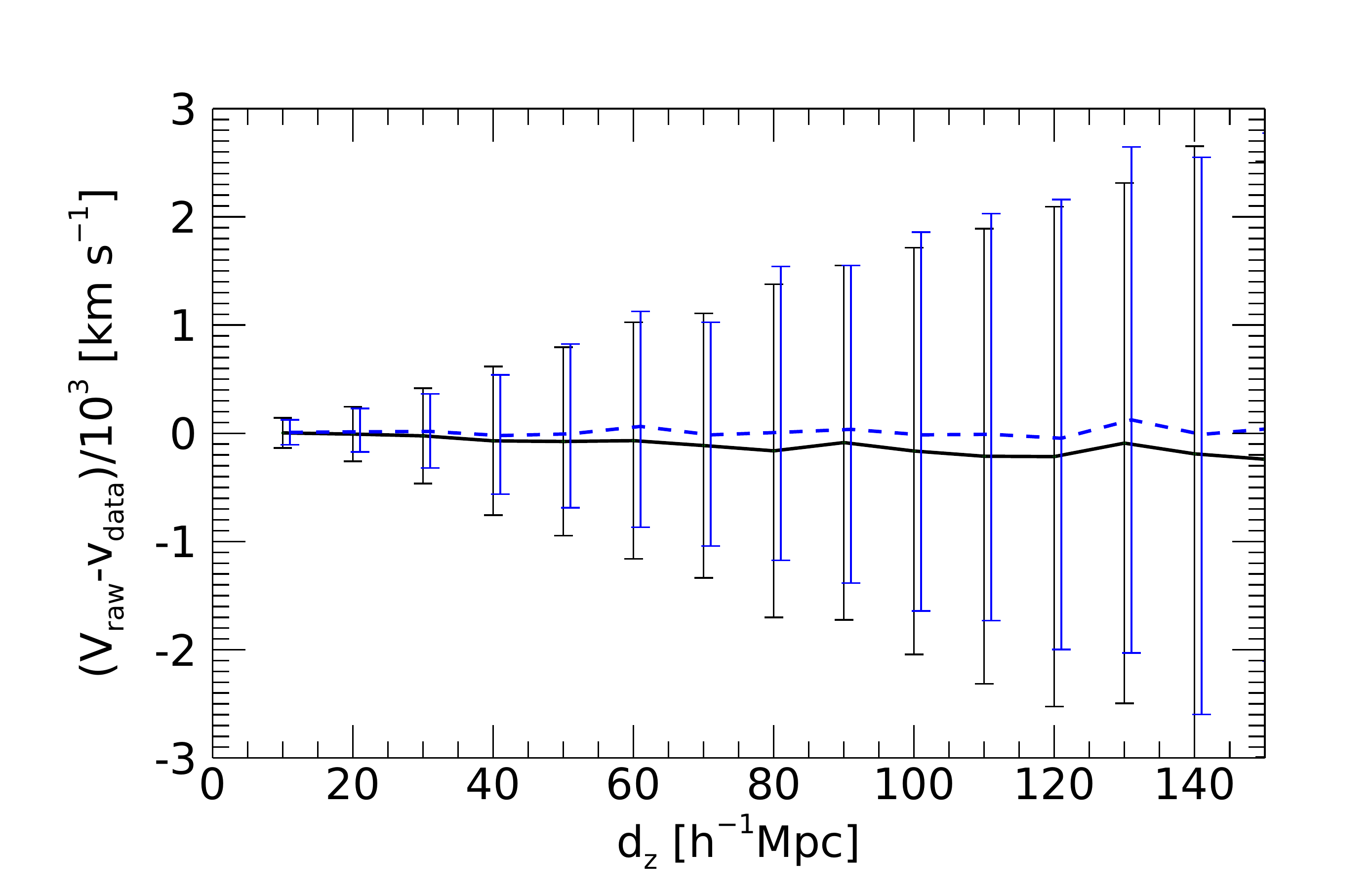}
}
\centerline{
\includegraphics[width=1.0\columnwidth]{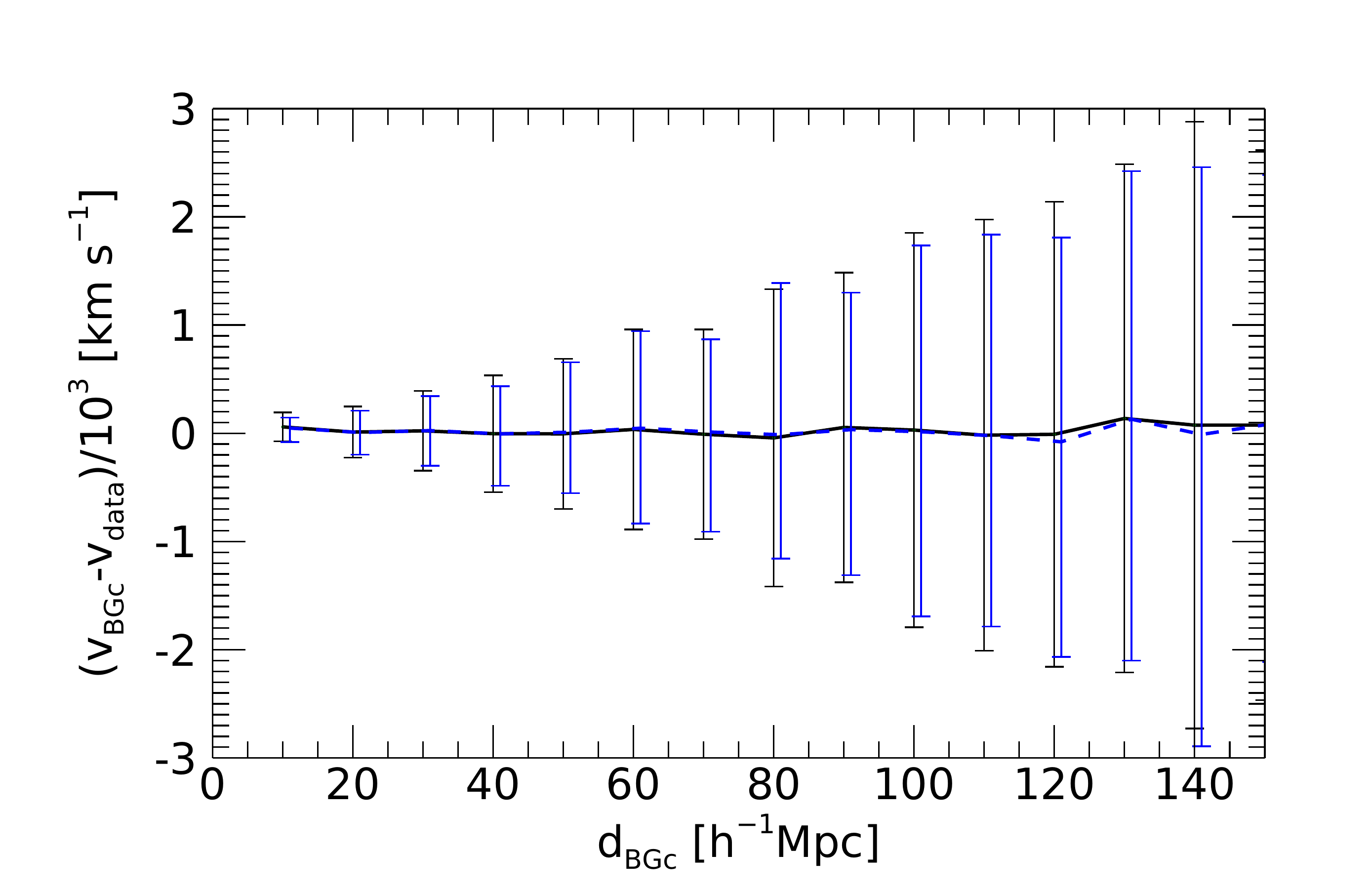}
}
\caption{ 
Mean (black, solid line) and median (blue, dashed line) 
of the   residual of the observed  velocity ($V$) from the true velocity  vs the observed distance ($D$, upper panel) 
and the residual of the estimated velocity ($v_{BGc}$) from the true data vs the estimated distance ($d_{BGc}$, lower panel). (Same notations as in Fig. \ref{fig:dBGc-D-residual-dz} .)
} 
\label{fig:Vrad-true-dz} 
\end{figure}

Fig. \ref{fig:dBGc-D-residual-dz} shows the residual of the observed distances from the true ones ($D-d$) vs their redshift distances (upper panel). The difference between the mean and the median of the distribution is a clear manifestation of the lognormal bias. 
The lower panel presents the residual of the BGc distances from the true ones. The mean and the median are virtually identical and they both oscillate around the null residual, as is expected from an unbiased Gaussian distance estimator.
Likewise,  Fig. \ref{fig:Vrad-true-dz}  shows the residual of the observed from the true velocities ($V-v$) and of ($v_{BGc}-v$ ) against $d_z$, and of the BGc estimated from the true velocities ($v_{BGc}-v$ ) vs $d_{BGc}$. Again, the later case is consistent with an unbiased  Gaussian distribution. 
There is one striking difference between the BGc correction of the distances and of the velocities. The BGc corrected distances are assumed to be normally scattered around $(D\vert z)_{med}$ with a constant scatter of 
 $\sim\sigma_{vp}/H_0\sim 2.75\hmpc$. 
In contrast the BGc velocities are assumed to be normally scattered around their corresponding median with the  estimated error  of the uncorrected data,  $\sigma_V\sim c z \tilde{\sigma}_\mu$.

\begin{figure}
\centerline{
\includegraphics[width=1.0\columnwidth]{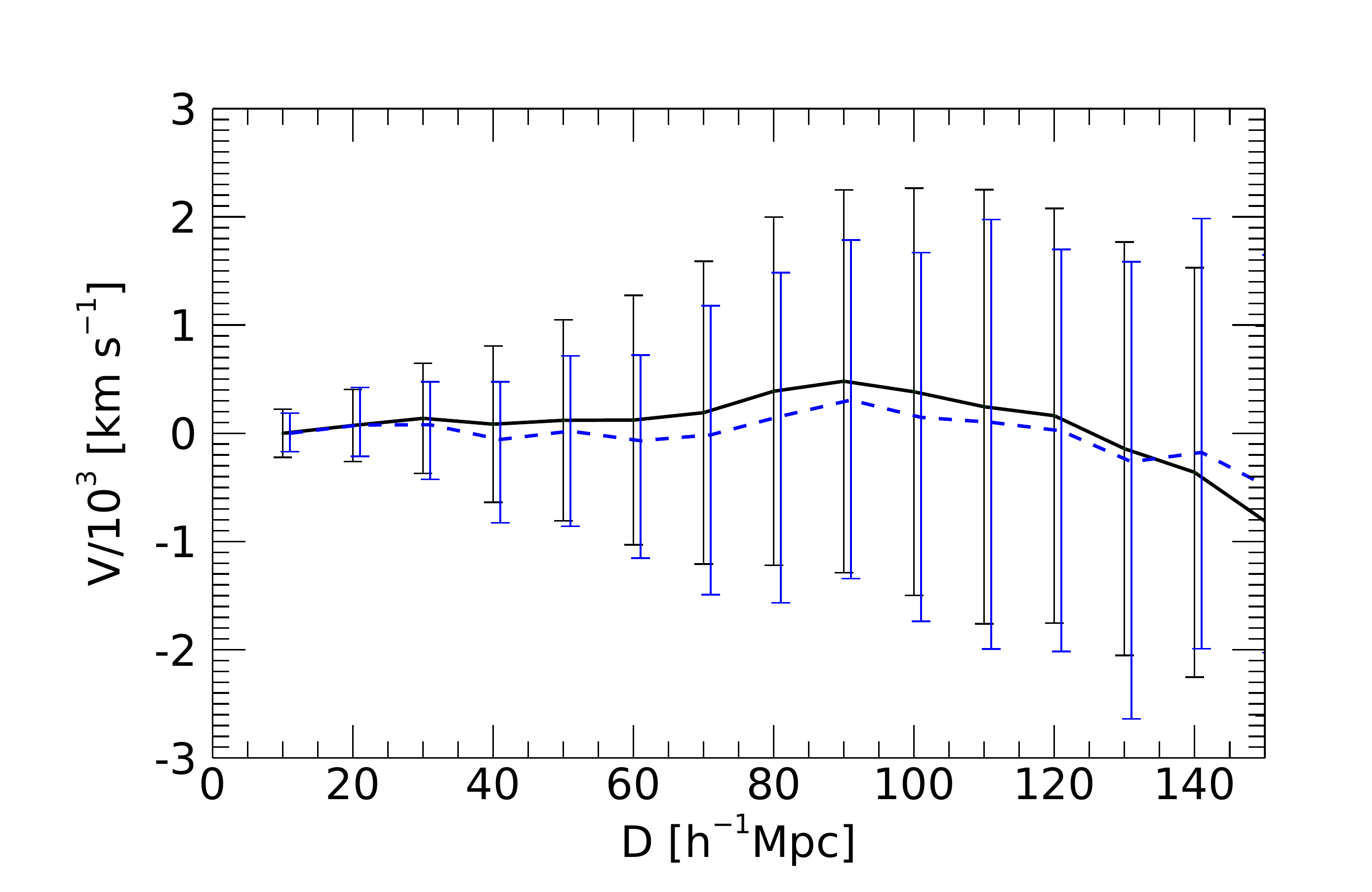}
}
\centerline{
\includegraphics[width=1.0\columnwidth]{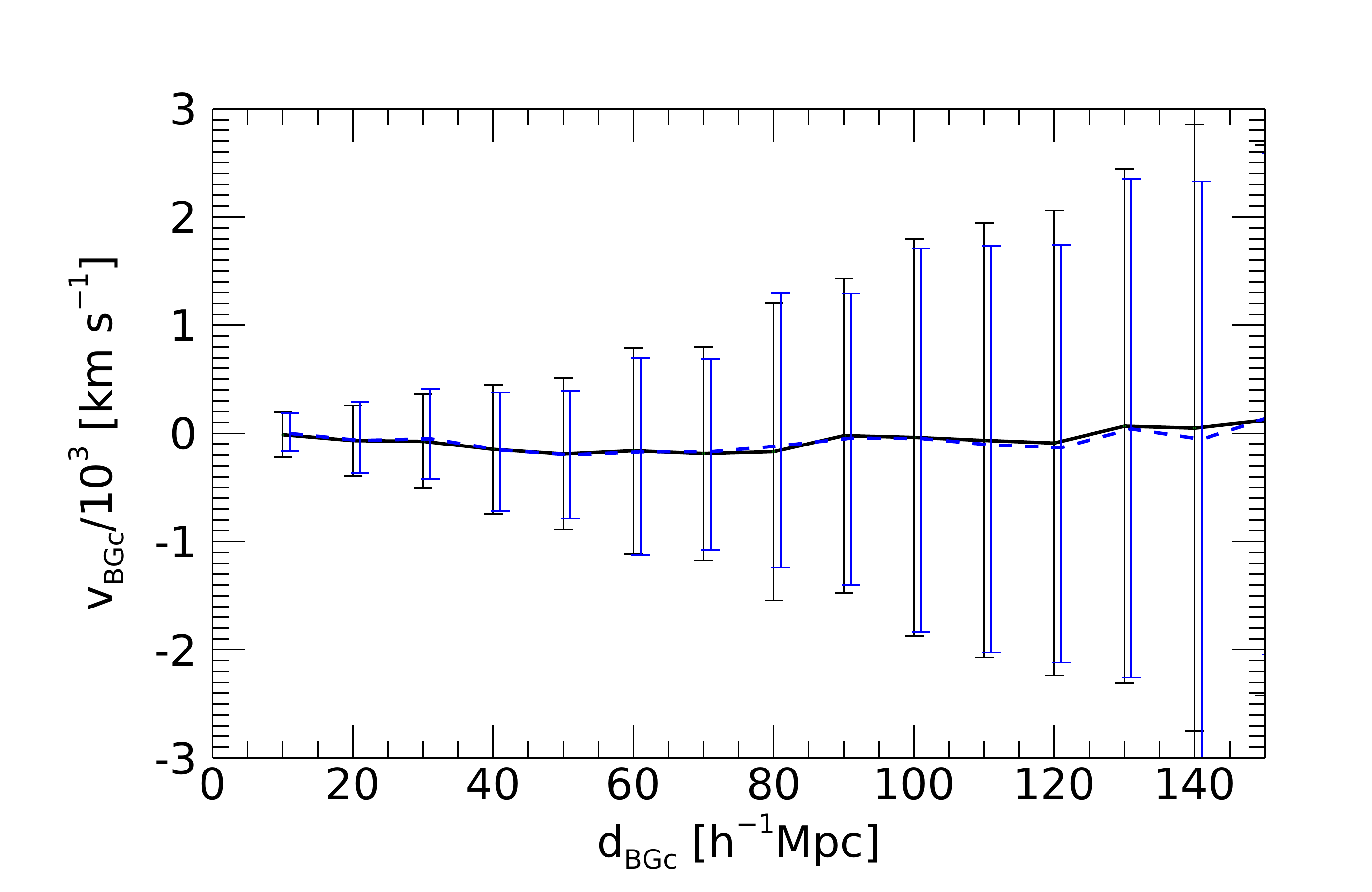}
}
\caption{ 
Mean (black, solid line) and median (blue, dashed line) 
of the   observed  velocity ($V$) vs the observed distance ($D$, upper panel) 
and the estimated velocity ($v_{BGc}$) vs the estimated distance ($d_{BGc}$, lower panel). (Same notations as in Fig. \ref{fig:dBGc-D-residual-dz} .)
} 
\label{fig:Vrad-dz} 
\end{figure}

 Fig. \ref{fig:Vrad-dz} presents the distribution of the observed velocities vs their observed distances (upper panel). The lognormal bias of the distances and velocities is clearly manifested here - the mean and median are different and both deviate significantly from zero. A naive reconstruction from such a data would show  strong false outflow (out to $D \sim 100 \hmpc$) and  inflow (beyond it). 
 The lower panel presents the scatter of the BGc velocities vs the BGc distances. The mean and the median are effectively identical and  both oscillate around zero, as expected from a Gaussian random field.

\section{Wiener filter reconstruction: comparison}
\label{sec:WF}

The aim here  is  to test the BGc estimation within the context   of the WF  reconstruction. 
The WF is the optimal estimator in the case where the data is drawn from an underlying random Gaussian field and observational `errors' that are normally distributed  \citep{1995ApJ...449..446Z}.  The case of  the grouped Cosmicflows data is challenging for the WF - distances  and velocities are subjected 
to lognormal uncertainties, yet the underlying velocity field is `almost' Gaussian. Accordingly,  the following  `ideal' mock data is constructed here - one in which the distances are known exactly and the `observed' velocities are  subjected to Gaussian errors given by Eq.   \ref{eq:eps-v} to generate the errors, with the same realization of the normal distribution used to generate the observed distance moduli. We call it here the $V_{Gauss}$ data and it is the aim of the BGc algorithm to recover it from the lognormal data. The BGc is tested here by comparing the WF reconstruction from the $v_{BGc}$ with the target WF  from the $V_{Gauss}$. A good  agreement  among the two reconstructions  means that the BGc indeed removes the lognormal bias successfully. The actual velocity field of the simulation is presented here as well. It is evaluated here by means of a clouds-in-cells (CIC) interpolation applied to the particles distribution over a grid of $128^3$ cells covering a box of side length of $500\hmpc$. The CIC velocity field is further Gaussian smoothed with a $5\hmpc$ Gaussian kernel, so as to suppress some of the non-linear components  of the velocity and to enable a comparison with the linear WF reconstruction. The nonuniform and anisotropic sampling of the data and the observational uncertainties that increase with distance implies that even for the $V_{gauss}$ case the quality of the WF reconstruction is expected to degrade with distance.

\begin{figure*}
\vskip -0.2cm
\centerline{
\includegraphics[width=0.43\textwidth]{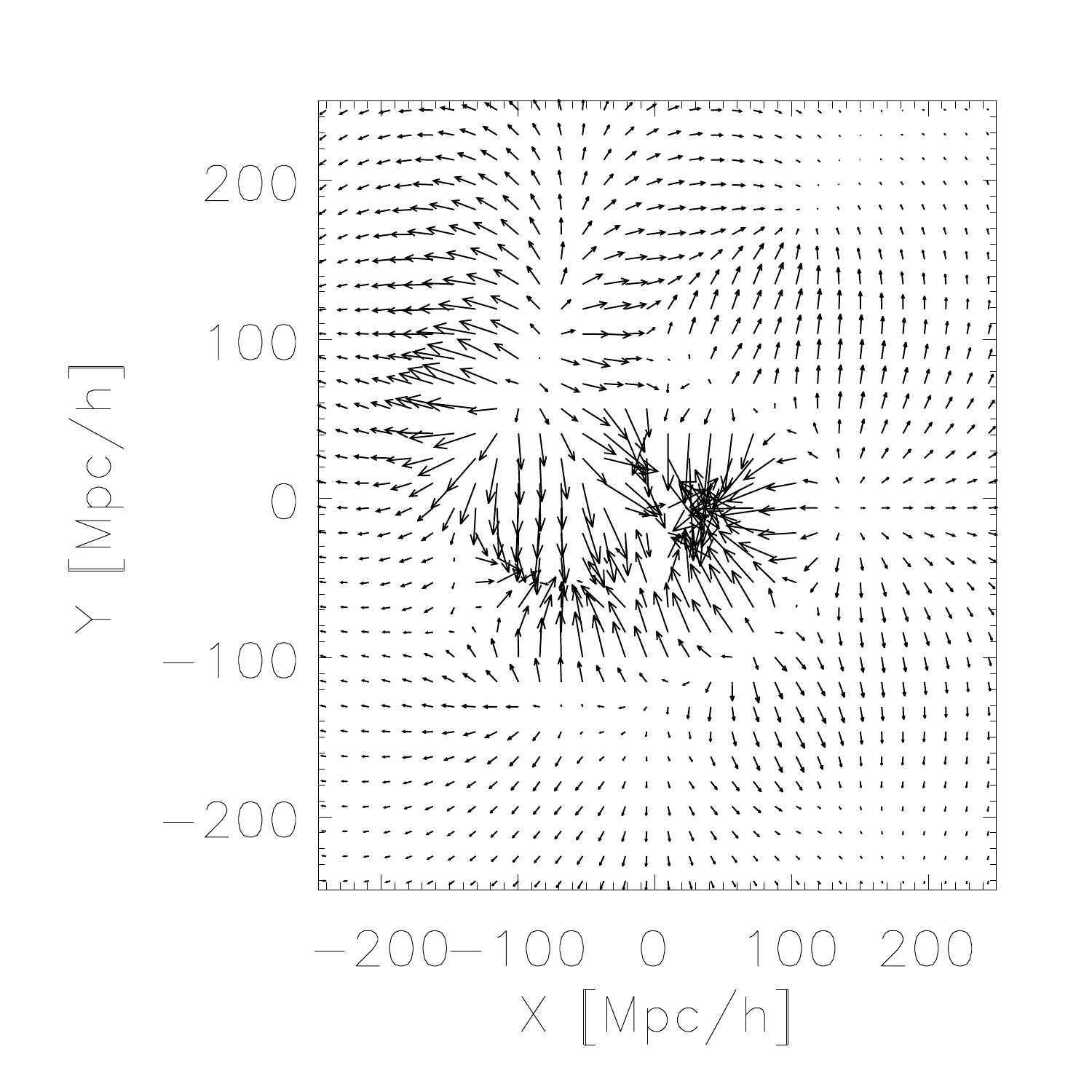}
\includegraphics[width=0.43\textwidth]{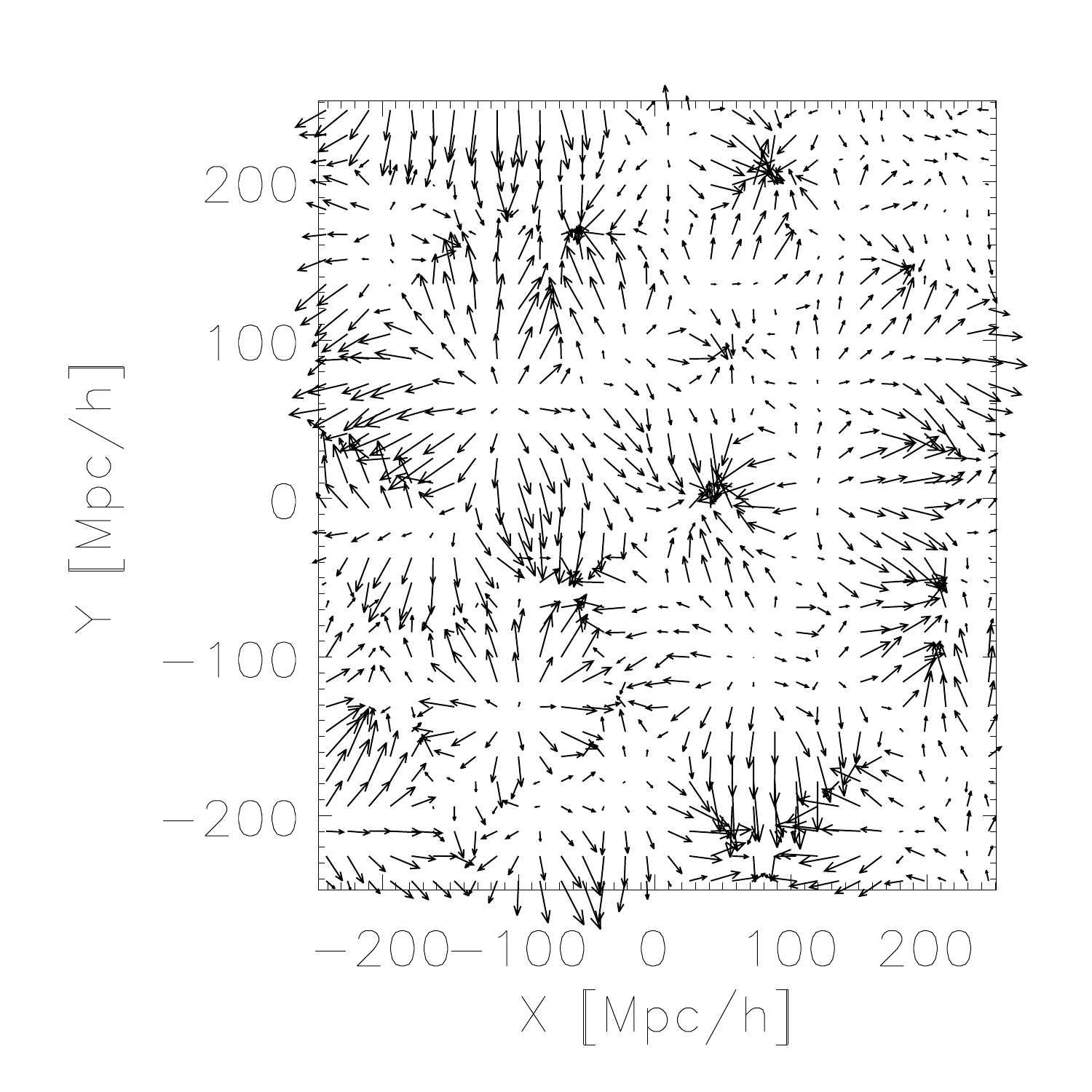}
}
\vskip -0.5cm
\centerline{
\includegraphics[width=0.43\textwidth]{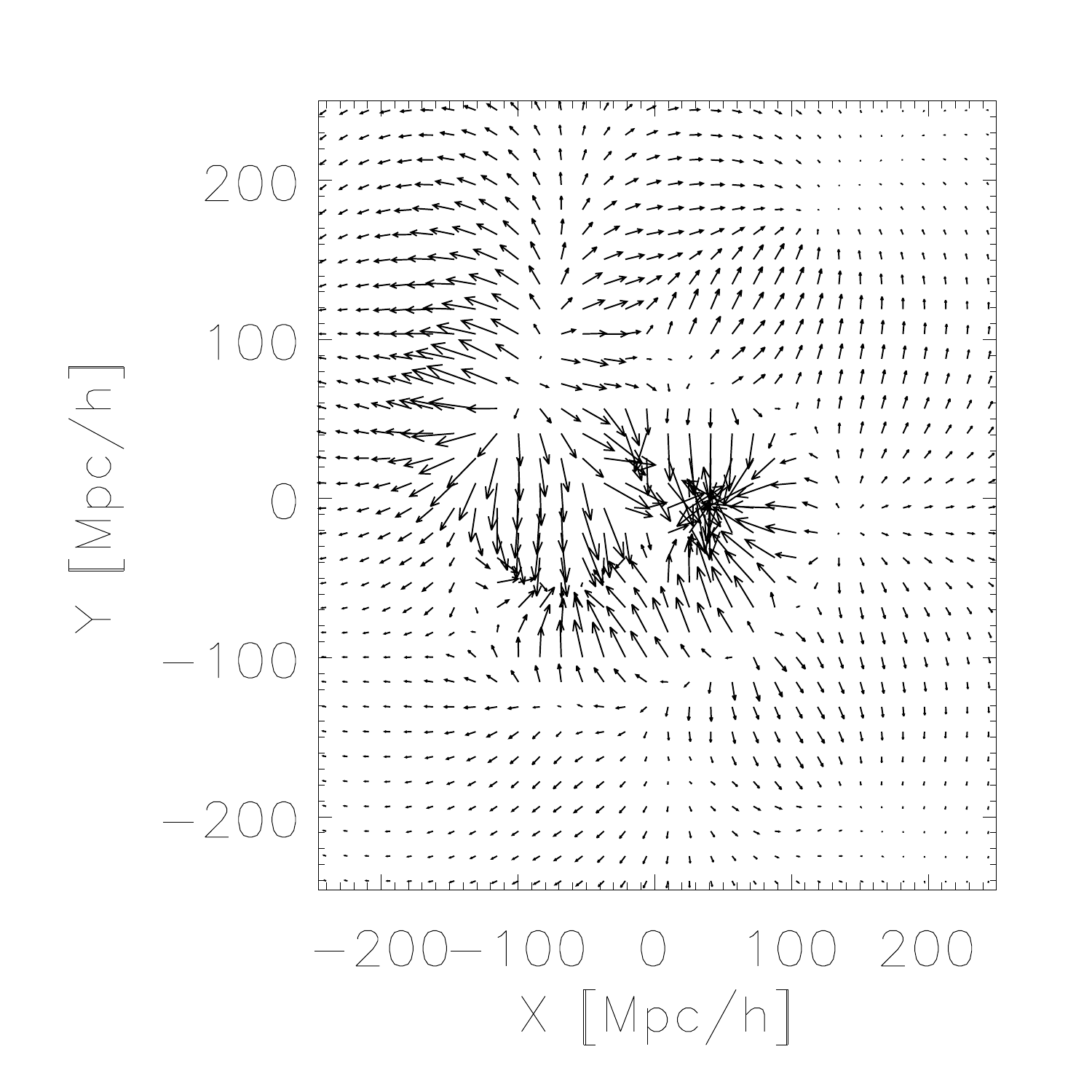}
\includegraphics[width=0.43\textwidth]{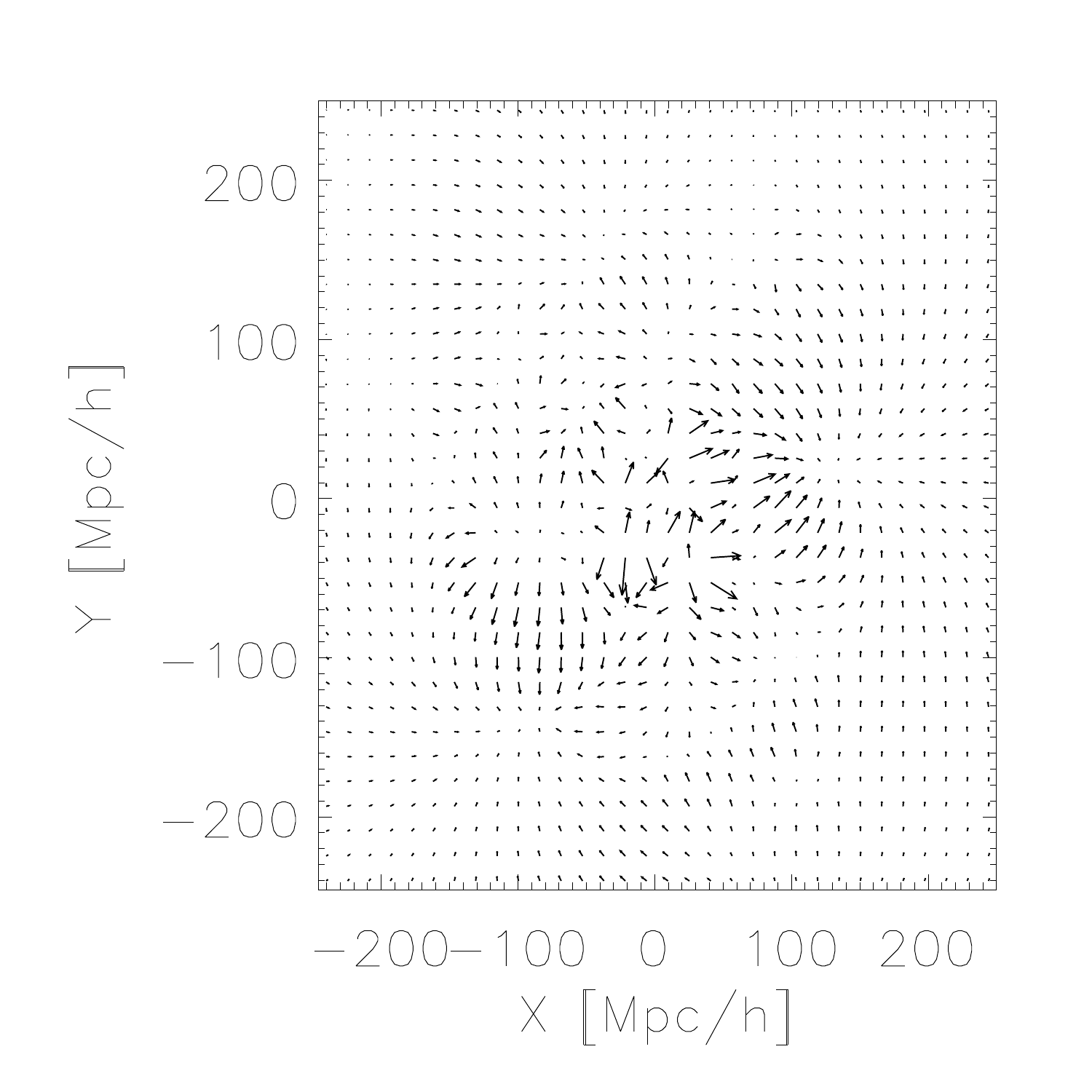}
}
\caption{ 
Visualization of flow fields:
a. The target WF reconstructed from the $V_{Gauss}$  data (see text) that is used here as a benchmark (upper-left panel) and the CIC velocity field, Gaussian smoothed with a $5\hmpc$  kernel (upper-right panel, the  velocity arrows length is  scaled down by a factor of 2 of the one used in the WF cases ). 
b. The WF reconstruction from the BGc estimated data (lower-left panel) and its residual of the WF reconstruction from the $V_{Gauss}$  (lower-right panel).
 The flow fields are evaluated at the mock Supergalactic Plane.
} 
\label{fig:dif-VGauss} 
\end{figure*}

Fig. \ref{fig:dif-VGauss}  depicts the WF reconstruction from the $V_{Gauss}$ data and the Gaussian smoothed CIC flow field. It further shows the WF reconstruction from the BGc estimated data and its residuals   from the target  WF/$V_{Gauss}$ field. The mock Supergalactic Plane of the velocity fields is shown here. 
This is relevant for the distribution of the data points, which follows that of the actual CF3 data, but is inconsequential  as far as the target simulation is concerned.
Note the different arrow lengths for the magnitudes of the streamlines of  WF/$v_{BGc}$  and its residual. The ratio of the mean  of the norm of the residual velocity field to the mean of norm of the velocity of the target field ranges from $\sim 0.25$ to $\sim 0.20$ within    $R=50$ and $150\hmpc$ (respectively). Visual inspection of the residual velocity field does not find any  clear coherence between the residual and target fields within the plane and in any case the magnitude of the residual field gets smaller by more than 20\% at large  distances. This is further inspected by evaluating the  multipole moments of the velocity fields.

\begin{figure}
\centerline{
\includegraphics[width=1.0\columnwidth]{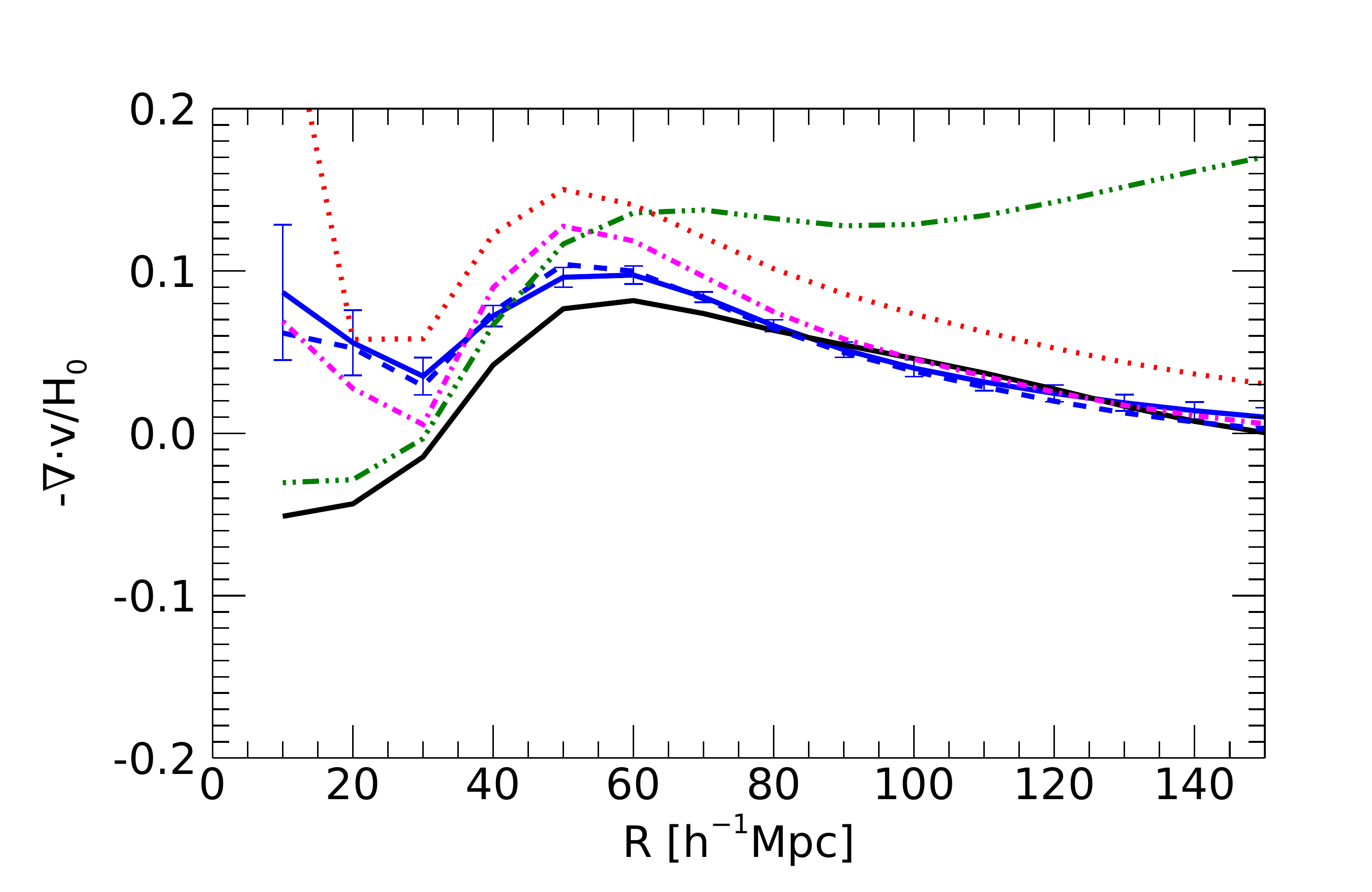}
}
\centerline{
\includegraphics[width=1.0\columnwidth]{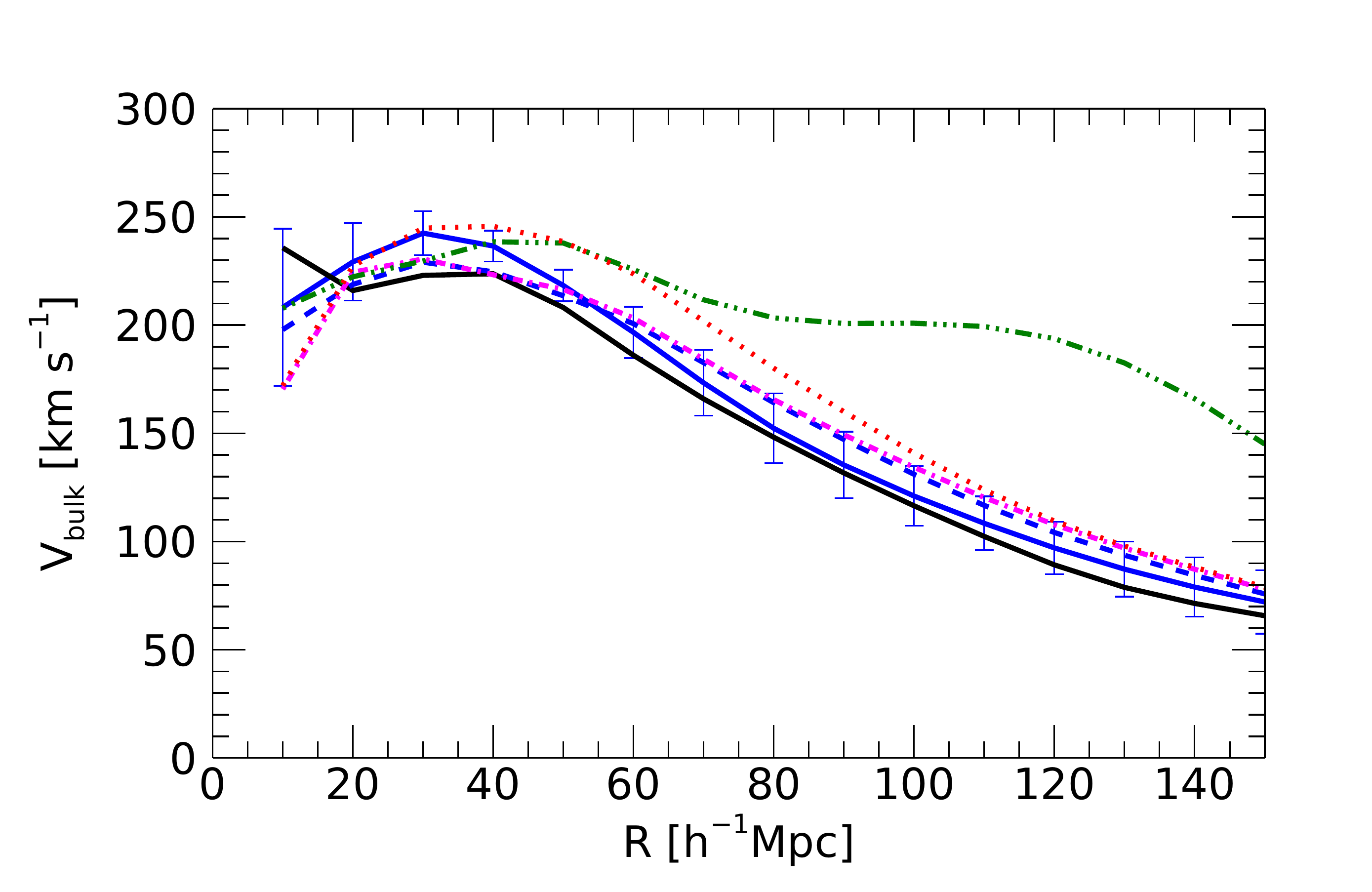}
}
\caption{ 
The monopole (upper panel) and the dipole  (namely the bulk velocity, lower panel) moments of the velocity field of the:
a. CIC smoothed velocity field of the simulation (black);
b The WF reconstruction from the  BGc corrected  data of  a single errors realization (dashed  blue);
c. The mean and the standard deviation of of the  WF   reconstruction from the  an ensemble of BGc data of 10 errors realizations (solid, blue);
d. The WF reconstruction from the    data with exact distances and  Gaussian errors on the velocities ($V_{Gauss}$)  of  the same errors realization (dot-dashed  magenta);
e. The WF reconstruction from the `observed' data (dot-dot-dashed green).
f. The WF reconstruction from the redshift distance and the `observed' uncorrected data (doted, red).
The main lesson to take here is the gross discrepancy between the multipoles of WF of the uncorrected data and the CIC and the CIC `true' field beyond $\approx50\hmpc$. This stands in sharp contrast  with the good agreement for the  WF of the BGc corrected data and the CIC cases.
} 
\label{fig:divv-dipole} 
\end{figure}

A further insight into the large scale behaviour of the WF reconstruction is provided by the volume-weighted mean monopole and dipole moments of the velocity field in spheres of radius $R$.
The monopole moment is  the  mean of    $-\nabla\cdot\vec{v}/H_0$, where the  scaling by $H_0$ is introduced so as to make the expression dimensionless and proportional to the mean (linear) over-density within $R$. The minus  sign is introduced so as to make   the monopole within a sphere of $R$  to be proportional to the mean  over-density within  that volume.
It corresponds  also to the   local fractional perturbation to $H_0$. The dipole  is the volume weighted mean velocity within $R$, i.e. the bulk velocity $V_{bulk}(R)$. 
Fig. \ref{fig:divv-dipole} presents the monopole and dipole moments, as a function of $R$, of the CIC Gaussian smoothed field and the WF reconstruction of the target data, of the BGc estimated data and of the  observed data. In addition, the mean and the standard deviation around the mean of the moments of the BGc estimators of an ensemble of 10 errors realizations are shown as well.  The monopole moment of the CIC and of the target data are robustly recovered for depth exceeding $\sim70\hmpc$. At smaller distances some deviations are found between the reconstructions from the target data and the BGc estimated data, and between these and the underlying CIC field. The dipole moment, namely the bulk velocity, is faithfully recovered across $10\ \lesssim\ R\  \lesssim \ 150\  \hmpc$. 

\begin{figure}
\centerline{
\includegraphics[width=1.0\columnwidth]{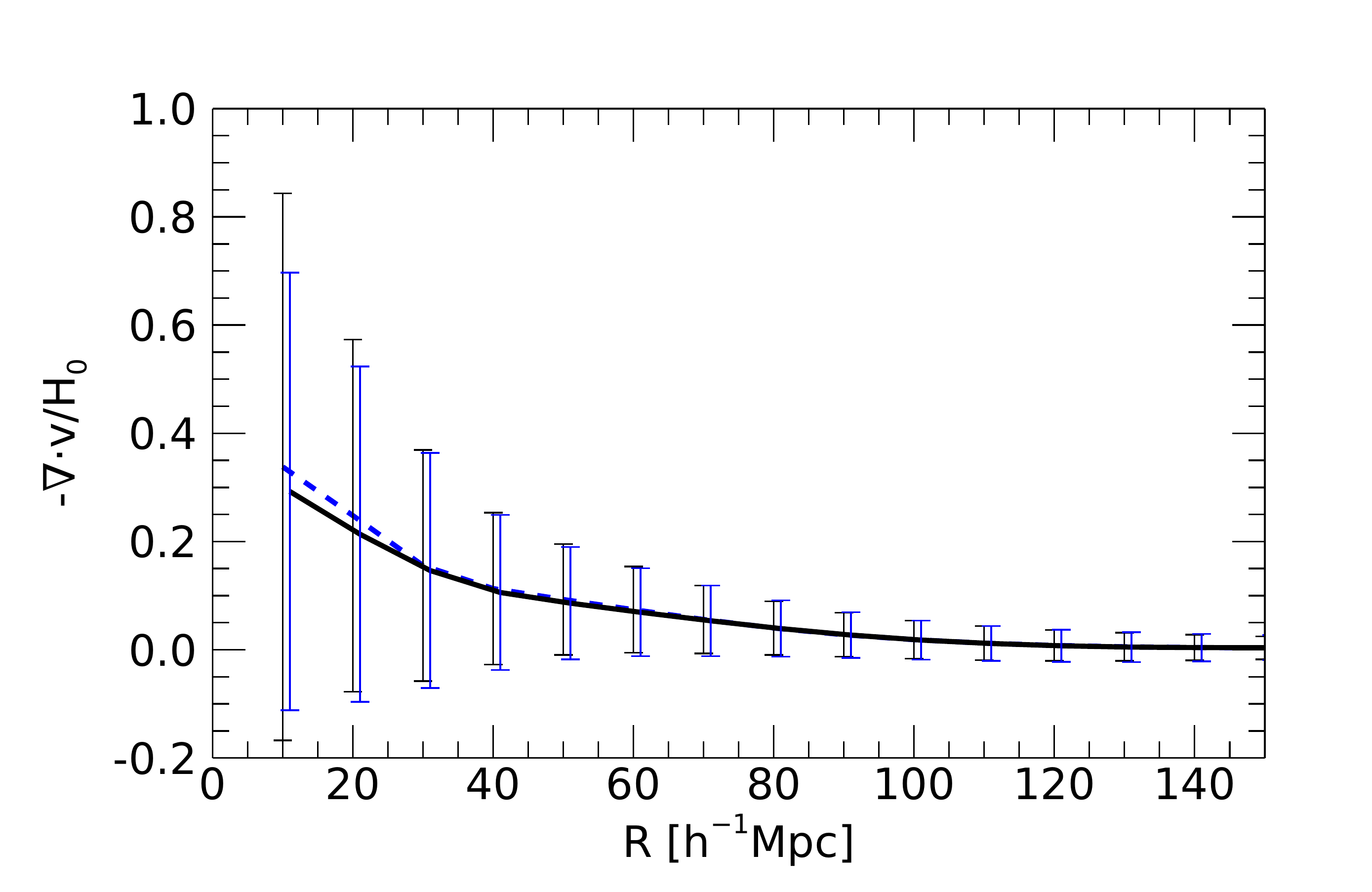}
}
\centerline{
\includegraphics[width=1.0\columnwidth]{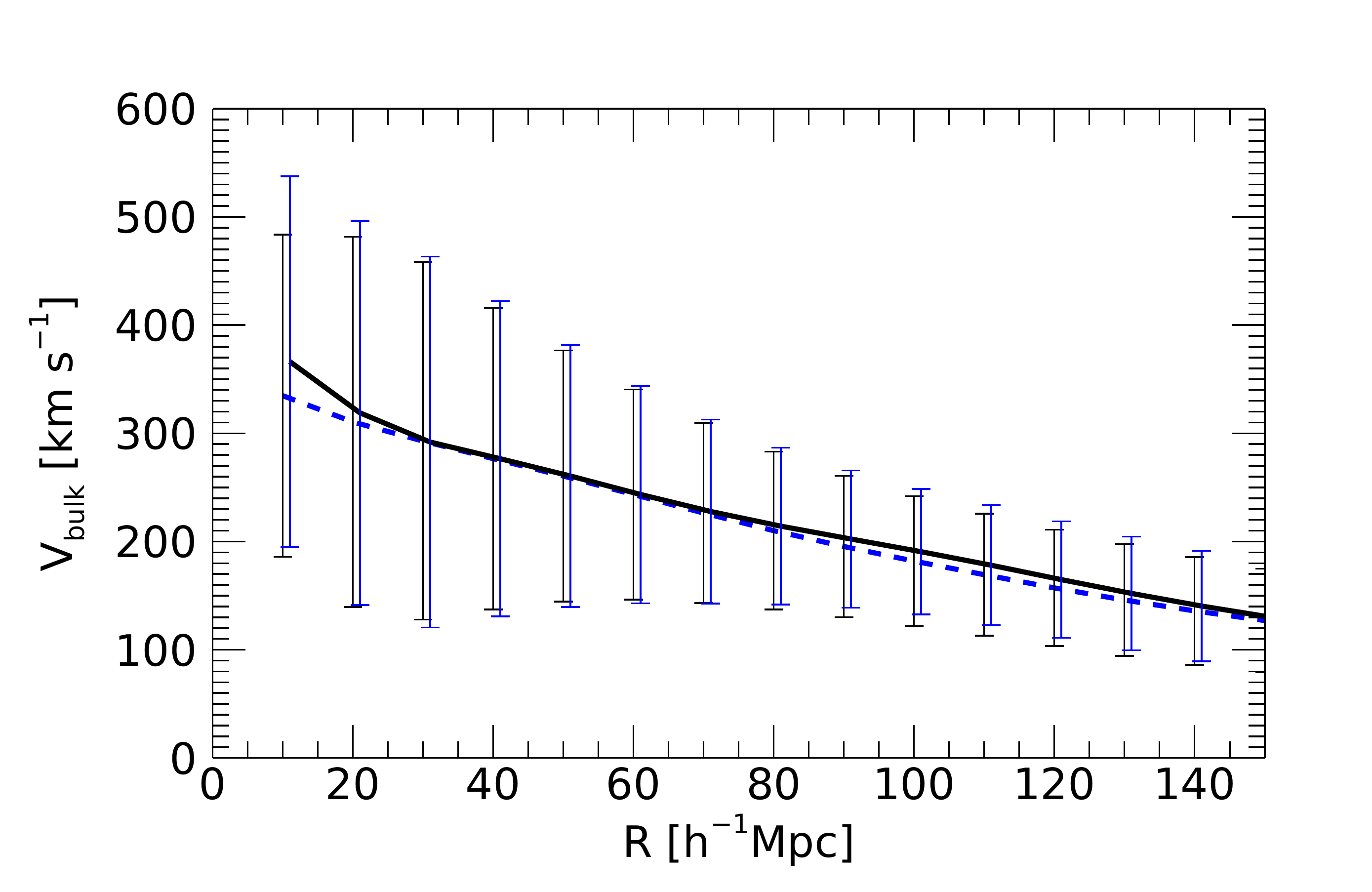}
}
\caption{ 
  The mean and scatter around the mean of the 
monopole (upper panel) and the dipole  (namely the bulk velocity, lower panel) moments of:
a. The CIC smoothed velocity fields of the ensemble of mock observers (cosmic variance; black). 
b. The   WF reconstructed velocity fields from the ensemble of 100 mock catalogs (cosmic and error variance, blue).
} 
\label{fig:CIC-BGc-stat_divv-dipole} 
\end{figure}

Fig. \ref{fig:CIC-BGc-stat_divv-dipole}  shows the cosmic variance around the 10 different mock observers as well as the errors variance, so as to gain further insight into the estimation of the monopole and dipole moments   . It shows  the mean and scatter of the moments taken over the CIC fields around the mock observers and of the BGc data of the 10 different errors realizations around the 10 different observers. For both moments the cosmic + errors variance of the WF/BGc fields agrees virtually perfectly with the cosmic variance over the CIC fields. Comparison Figs. \ref{fig:divv-dipole}  and \ref{fig:CIC-BGc-stat_divv-dipole} shows that the cosmic variance dwarfs the variance due to the `observational' errors.  

The monopole moment discrepancy between the CIC, the target WF field and the WF/BGc fields (Fig. \ref{fig:divv-dipole}) can be partially attributed to the inherent nonlinearity of the CIC field.
The approximate  agreement between  the target field and the WF/BGc reconstructions  suggests that the BGc is doing well in removing the  lognormal bias. The moments of the WF  of the uncorrected `observed' velocities are clearly  misrepresenting the underlying field for $ R\  \gtrsim \ 50\  \hmpc$.

\section{A closer look at the estimation of the velocities }
\label{sec:other-schemes}

The mathematical expressions for the BGc estimated distances and velocities, Eqs. \ref{eq:dg} and \ref{eq:vg} consist of two terms - the first is the median of the relevant lognormal distribution and the other is its  Gaussianization. Centralizing the resulting normal distribution on the median removes the biased mean value introduced by the lognormal transformation 
(Eqs. 11 and 25)
It is interesting to separate these two ingredients of the BGc procedure and check how they affect the WF reconstruction.

\begin{figure}
\centerline{
\includegraphics[width=1.0\columnwidth]{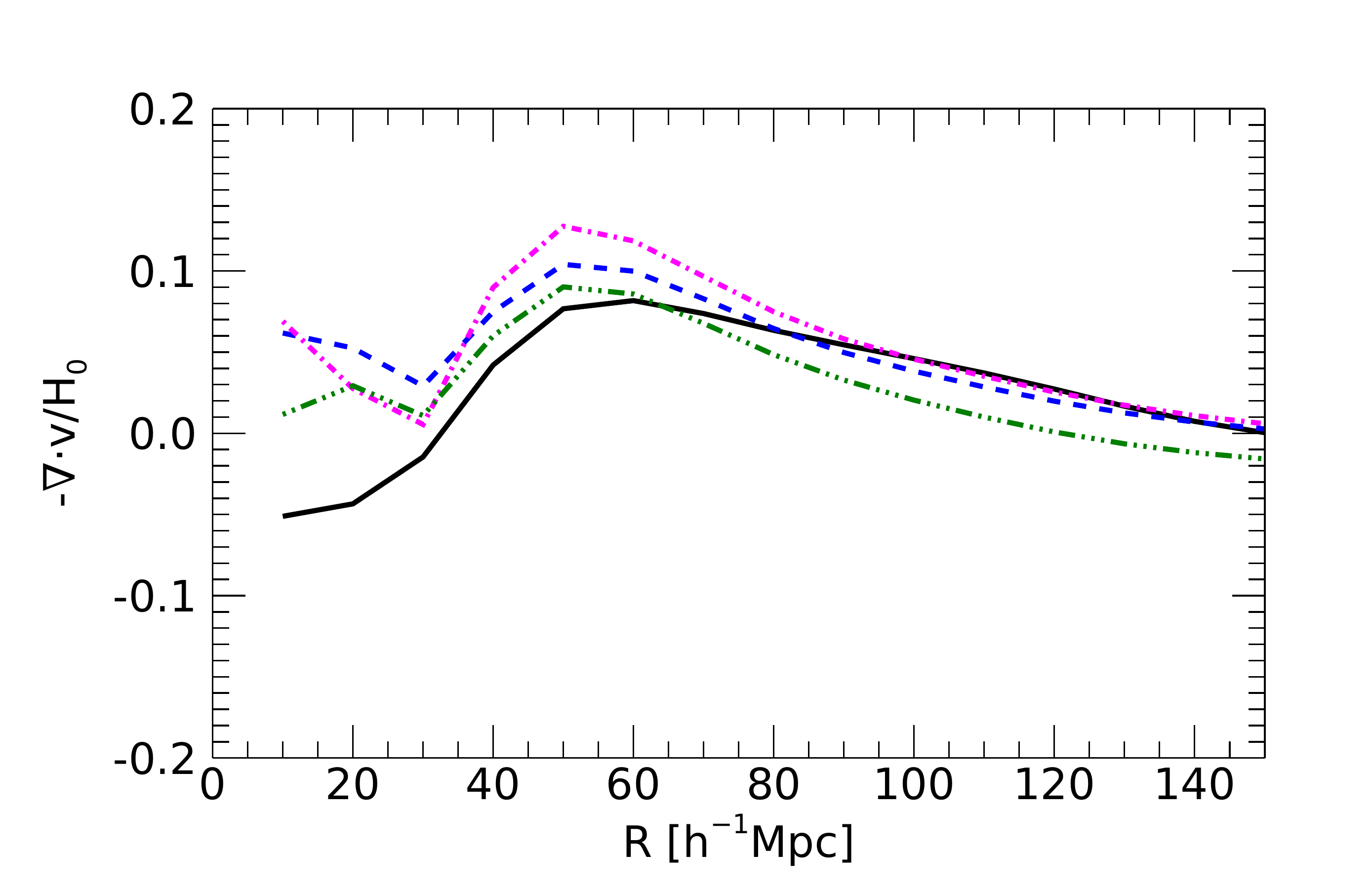}
}
\centerline{
\includegraphics[width=1.0\columnwidth]{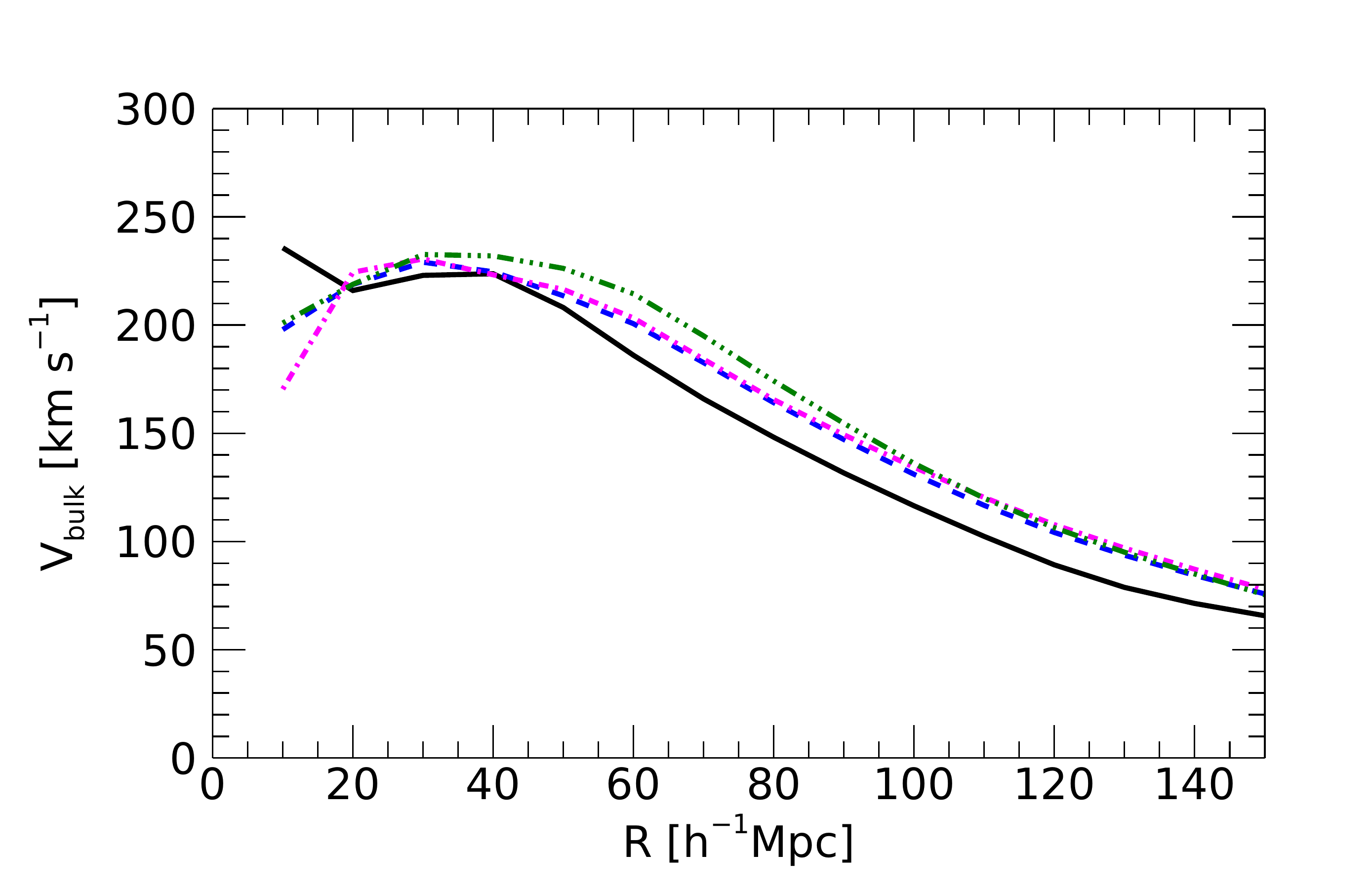}
}
\caption{ 
The monopole (upper panel) and the dipole  (lower panel) moments of the WF reconstructed velocity fields from the following data sets: 
a. BGc (blue, dashed); 
b. The VGauss target ideal data (magenta, dot-dashed); 
c. BGc distances ($d_{BGc}$) and observed velocities shifted by the lognormal bias (green, dot-dot-dashed);  
The moments of the Gaussian smoothed CIC velocity field is presented for reference (black, solid).
} 
\label{fig:compare_divv-dipole} 
\end{figure}

This is done here by constructing from  a data set made of the BGc estimated velocities and velocities defined by 
\begin{equation}
v_{m-shift}=(V\vert z)_{\rm med}.
 \label{eq:v-shift}
\end{equation}
Namely, the Gaussianization term is omitted from Eq. \ref{eq:vg} and the estimated velocities, $v_{m-shift}$, are set to the median of the lognormal distribution. Fig. \ref{fig:compare_divv-dipole} depicts the monopole and dipole moments of the WF reconstruction of the BGC estimated distances and velocities ad of the $v_{m-shift}$ data. The cases of the the WF reconstruction from the ideal $V_{Gauss}$ target data and Gaussian smoothed ($R_s=5\hmpc$ kernel) CIC are shown for reference. The dipole moment is less sensitive to the different data sets used here compared with the monopole moment. The dipole moments of the BGc (distances and velocities) and the $V_{Gauss}$ data sets are virtually identical (for $R\geq20\hmpc$) all the way to the edge of the data.The $v_{m-shift}$ case overestimate the bulk velocity of the target $V_{Gauss}$ by $\sim7\%$ at $R\sim 60 \hmpc$. As for the monopole moment, one finds very close agreement between the BGc, $V_{Gauss}$ and the CIC cases, while the  $v_{m-shift}$ exhibits an close to a  offset of roughly $-0.02$ for $R \gtrsim 60 \hmpc$. Such an offset corresponds to an overestimation of the value of Hubble's constant by roughly $1.5\kmsMpc$.

We conclude here that both ingredients in the BGc estimation of the velocities - the correction of the mean by the median and the Gaussianization - are essential for getting an unbiased estimator of the input velocities for the WF reconstruction.

\section{Estimation of the Hubble constant: mock catalogs}
\label{sec:H0-mocks}

The simulation was run within the framework  of the standard  \LCDM\  model and its cosmological parameters - in particular $H{^{\rm \Lambda CDM}_0}=67.7\kmsMpc$ was used. We set ourselves here to study the derived $H_0$   from each CF3-like mock catalogs, constructed to sample the cosmic and errors variances. 
Four distinct classes of estimated $H_0$ are used here:
{\bf a.} $H{^{\rm Nbody}_0}$ - the value estimated from the 'clean' data (i.e no errors added); 
{\bf b.}  $H{^{\rm obs}_0}$ - the value estimated from the observed data;
{\bf c.}  $H{^{\rm BGc}_0}$ - the value estimated from the BGc corrected data;
{\bf d.}  $H{^{\rm mean}_0}$ - the value estimated from the BGc corrected data, for which the Gaussianization of the distances is performed around the mean of the distribution. Namely, in Eq. \ref{eq:vg}  $(D\vert z)_{\rm med}$ is replaced by the mean distance of the data points of the $z$ subsample.The rational behind the    $H{^{\rm mean}_0}$ estimation is that the aim here is to overcome the scatter induced by the peculiar velocities around the observed redshift.

\begin{figure}
\centerline{
\includegraphics[width=1.0\columnwidth]{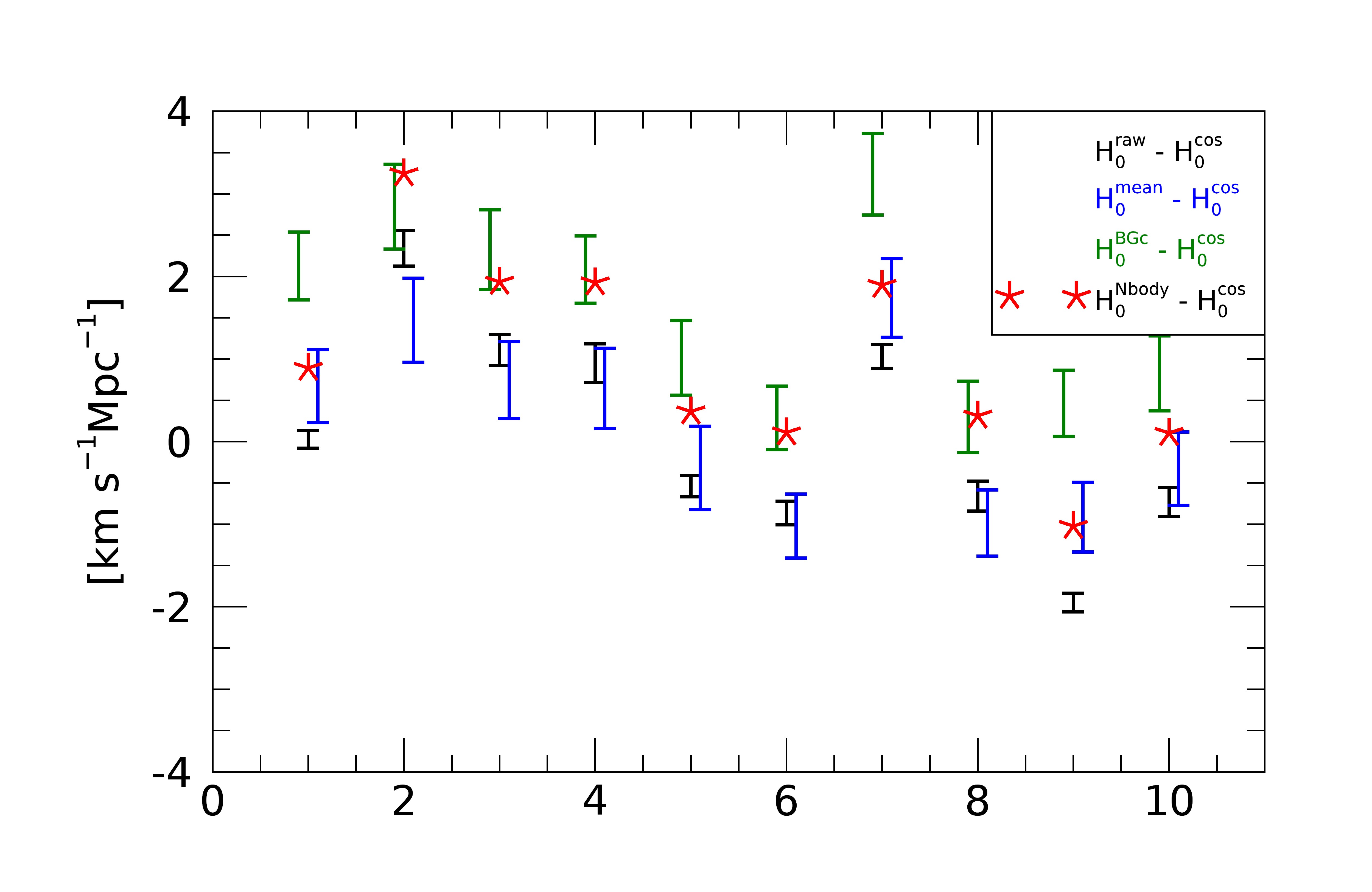} 
}
\caption{ 
Error variance: The mean and scatter of the residual of the estimated values of $H_0$ from the \LCDM\ assumed value  of $H{^{\rm \Lambda CDM}_0}=67.7\kmsMpc$ for each one of the mock CF3 observer (denoted by horizontal axis). Three case are considered here: $H{_0^{\rm obs}}$ (black), $H{_0^{\rm mean}}$ (blue) and $H{_0^{\rm BGc}}$ (green). The red asterisk symbols represent the  $H{_0^{\rm Nbody}}$ of the individual mock observers. The horizontal axis labels the different mock catalogs.
} 
\label{fig:H0-677-err-variance} 
\end{figure}

\begin{figure}
\centerline{
\includegraphics[width=1.0\columnwidth]{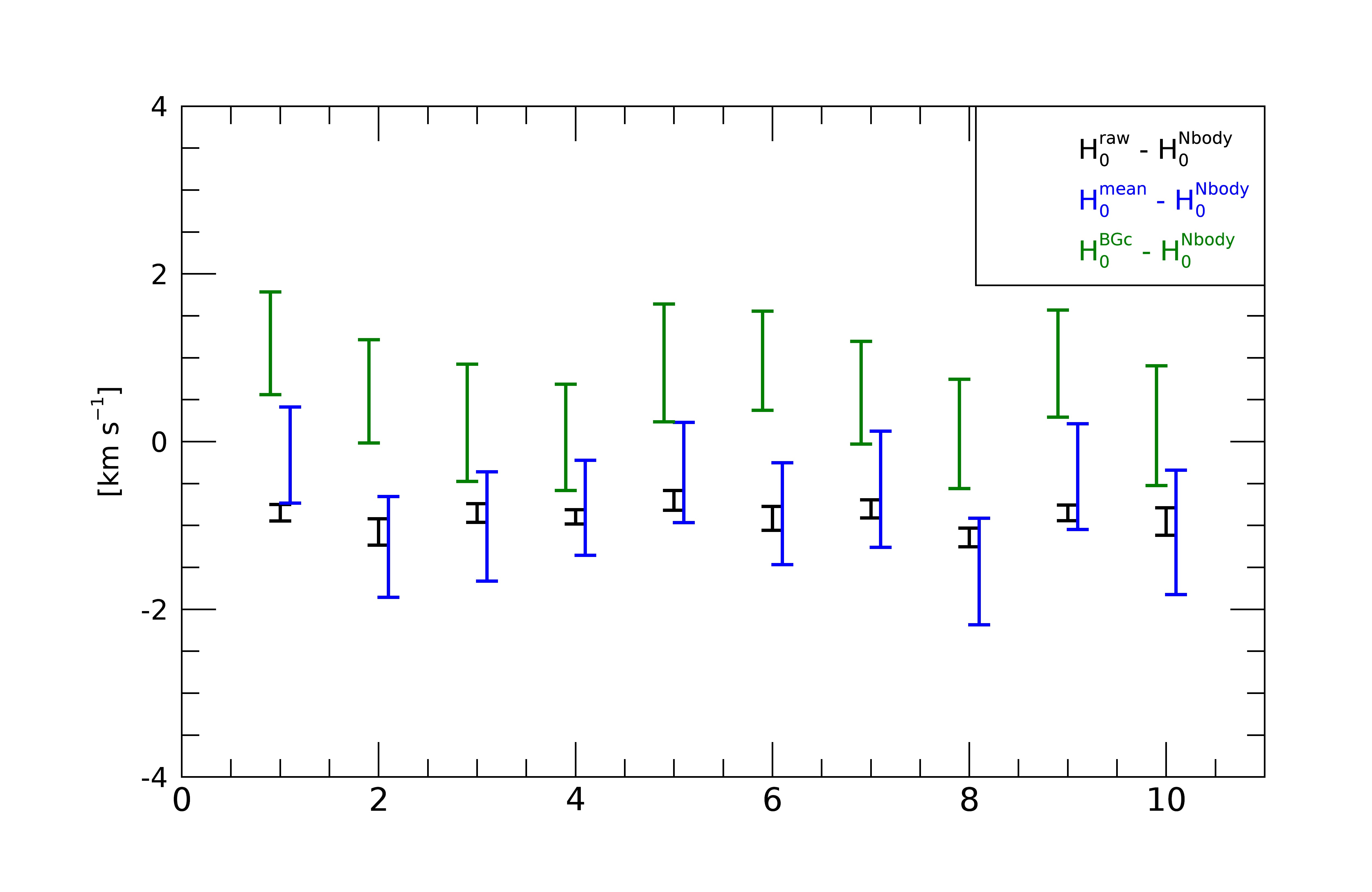} 
}
\caption{ 
Cosmic variance: The mean and scatter of the residual of the estimated values of $H_0$ from the actual $H{^{\rm Nbody}_0}$ for each one of the different errors realization (denoted by horizontal axis). Three case are considered here: $H{_0^{\rm obs}}$ (black), $H{_0^{\rm mean}}$ (blue) and $H{_0^{\rm BGc}}$ (green). The horizontal axis labels the different mock catalogs.
} 
\label{fig:H0-H0Nbody-cos-variance} 
\end{figure}

\begin{figure}
\centerline{
\includegraphics[width=01.1\columnwidth]{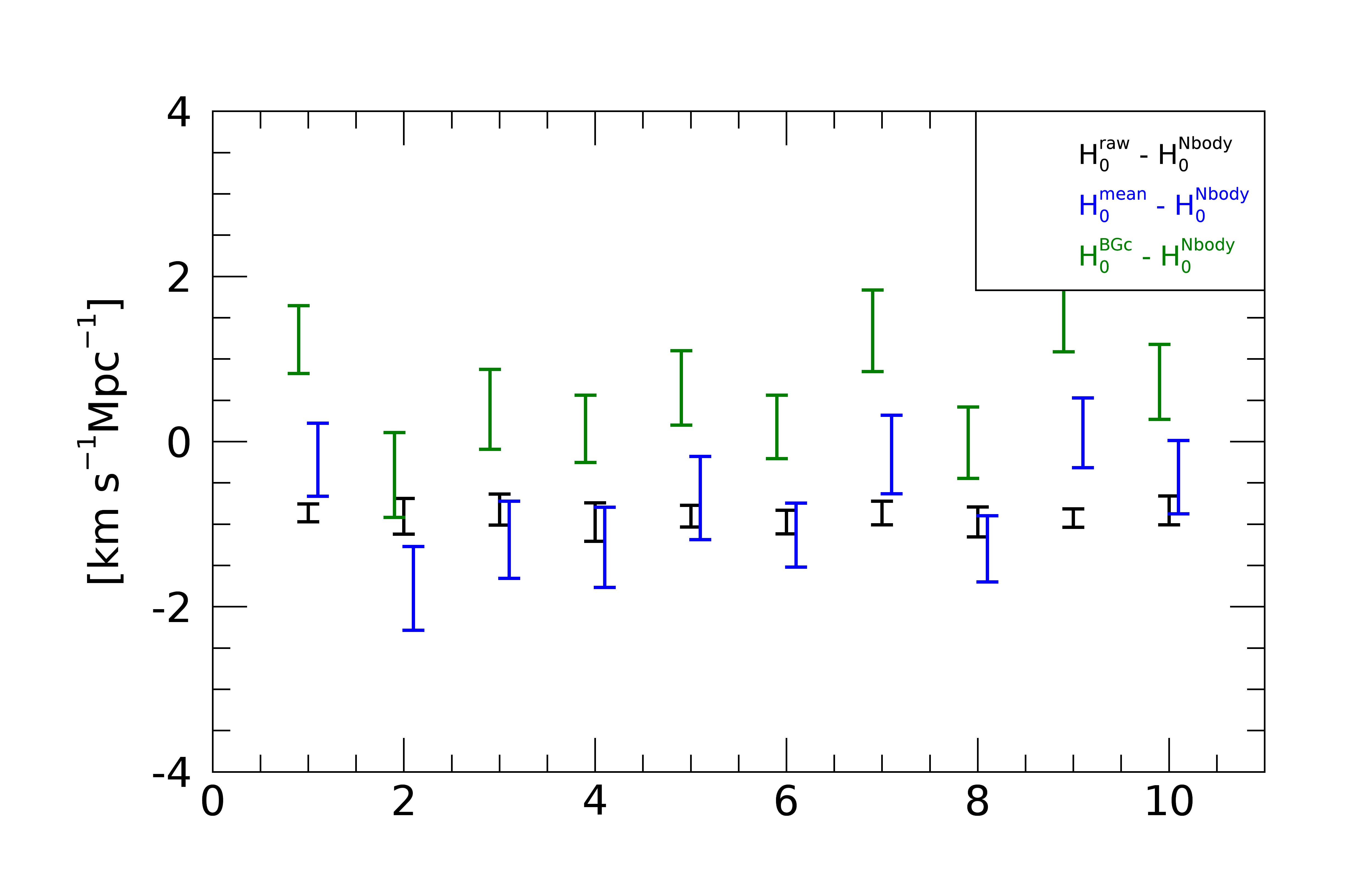} 
}
\caption{ 
Error variance: The mean and scatter of the residual of the estimated values of $H_0$ from the actual $H{^{\rm Nbody}_0}$ for each one of the mock CF3 observer (denoted by horizontal axis). Three case are considered here: $H{_0^{\rm obs}}$ (black), $H{_0^{\rm mean}}$ (blue) and $H{_0^{\rm BGc}}$ (green). The horizontal axis labels the different mock catalogs.
} 
\label{fig:H0-H0Nbody-err-variance}
\end{figure}

The Hubble constant is estimated here by linear regression over the distance-redshift scatter, in which both distances and redshift uncertainties are considered, and over all data points in the range of $20 \leq d_z \leq 150 \ \hmpc$. The lower limit of the distance range has been chosen to correspond to that  used for the case of the actual CF3 catalog, so as  to exclude the effect of the Virgo cluster.  The upper limit coincides with the range over which the BGc is applied. The formal statistical errors of the linear regression are very small in all case ($\lesssim 0.1\kmsMpc$) and are ignored here. The results are summarized in Table \ref{table:H0-mocks} and are visually presented in the following figures:
{\bf a.} The  error variance of the residual of the estimated $H_0$ from the assumed $H{^{\rm \Lambda CDM}_0}$ of the simulation  (Fig. \ref{fig:H0-677-err-variance});
{\bf b.} The cosmic variance of the residual of the estimated $H_0$ from $H{^{\rm Nbody}_0}$ (Fig. \ref{fig:H0-H0Nbody-cos-variance});
{\bf c.} The error variance of the residual of the estimated $H_0$ from $H{^{\rm Nbody}_0}$ (Fig. \ref{fig:H0-H0Nbody-err-variance} ).
 Here the estimated values of $H_0$ correspond to $H{^{\rm obs}_0}$, $H{^{\rm BGc}_0}$ and $H{^{\rm mean}_0}$. The cosmic variance refers to the variance with respect to the different mock  observers that are randomly scattered within the computational box of the simulation and the error variance corresponds to the scatter of the different realizations of the mock observational errors.
 
The fact that all the different estimators of $H_0$ yield an estimate that is within $\approx\pm1.0$ with a scatter of  $\approx\pm1.0\kmsMpc$ from $H{^{\rm \Lambda CDM}_0}$ is comforting.
 The finding that  $H{^{\rm obs}_0}-H{^{\rm \Lambda CDM}_0}=0.1\pm1.2\kmsMpc$, where the variance includes both the cosmic and the error variance, might seem surprising givin
 There is some similarity between the ways the monopole term of the velocity field  and the $H_0$ are constructed, hence 
 in light of the gross discrepancy between the monopole term of the uncorrected data (distances and velocities)  and the CIC (`true') one the lack of such discrepancy with regard to the different estimates of $H_0$ might appear surprising. But one should recall that the 'uncorrected' $H_0$ is derived essentially by linear regression of the uncorrected velocities on the the redshifts. The bias of the scatter of the velocities around the redshifts is much smaller than that of the velocities around the distances, as manifested by the dotted-red curve of Fig. \ref{fig:divv-dipole}. The curve shows that  at $R=150 \hmpc$ the mean value of the monopole term  is $\sim 3.0\%$,  namely $\nabla\cdot\vec{v}/H_0 \sim -0.03$. Assuming statistical anisotropy of the peculiar velocity field on the scale of the full CF3 data one expects the value of $H_0$ derived from that WF reconstructed velocity field to be deficient by roughly a third of that value, i.e.  $\sim1.0\%$. Indeed, from Table \ref{table:H0-mocks} one infers that  $(H{^{\rm obs}_0}  - H{^{\rm Nbody}_0} )/H{^{\rm Nbody}_0}  = -0.013 \pm 0.003$ - consistent with the WF reconstructed from the observed uncorrected velocities and the redshift distances.

\begin{table*}
\begin{tabular*}{\textwidth}{@{\extracolsep{\fill}}|c|c|c|c|c|c|c|}
\hline
 $H{^{\rm Nbody}_0}  - H_0$  &  $H{^{\rm obs}_0}  - H_0$   &   $ H{^{\rm mean}_0}  - H_0 $  &  $ H{^{\rm BGc}_0}  - H_0 $   &  $H{^{\rm obs}_0}  - H{^{\rm Nbody}_0} $  &  $H{^{\rm mean}_0}  - H{^{\rm Nbody}_0} $ &  $H{^{\rm BGc}_0}  - H{^{\rm Nbody}_0} $  \\
  \hline 
                                               &   $ 0.1 \pm 1.3 \pm 0.2 $      &  $0.2 \pm 1.0 \pm 0.4$             &  $ 1.6\pm 1.1\pm 0.4$            &  $-0.9 \pm 0.1\pm 0.2$                               & $-0.8\pm 0.5\pm 0.6$                                   & $0.6\pm0.6\pm0.4  $                                   \\
   \hline
 $1.0 \pm 1.2 $                      &      $0.1 \pm1.3 $                   &     $0.2\pm1.1  $                       & $ 1.6\pm1.2$                          &  $-0.9\pm0.2$                                             & $-0.8\pm0.8$                                                 & $0.6\pm0.7$                                                \\
\hline   
\end{tabular*}
\caption{The mean and standard deviation of the residual of $H{^{\rm obs}_0}$ and of  $H{^{\rm BGc}_0}$ from the 
actual $H{^{\rm \Lambda CDM}_0}$and from the actual value measure by each mock observer $H{^{\rm Nbody}_0} $: The upper observed presents separately the scatter due to the  cosmic variance (1st uncertainty) and the error variance (2nd uncertainty). The lower observed shows the total uncertainty. (All values are in units of $\kmsMpc$.)
}
\bigskip
\label{table:H0-mocks}
\end{table*}

\section{Cosmicflows-3 data: BGc}
\label{sec:CF3-BGc}
\begin{figure}
\centerline{
\includegraphics[width=1.0\columnwidth]{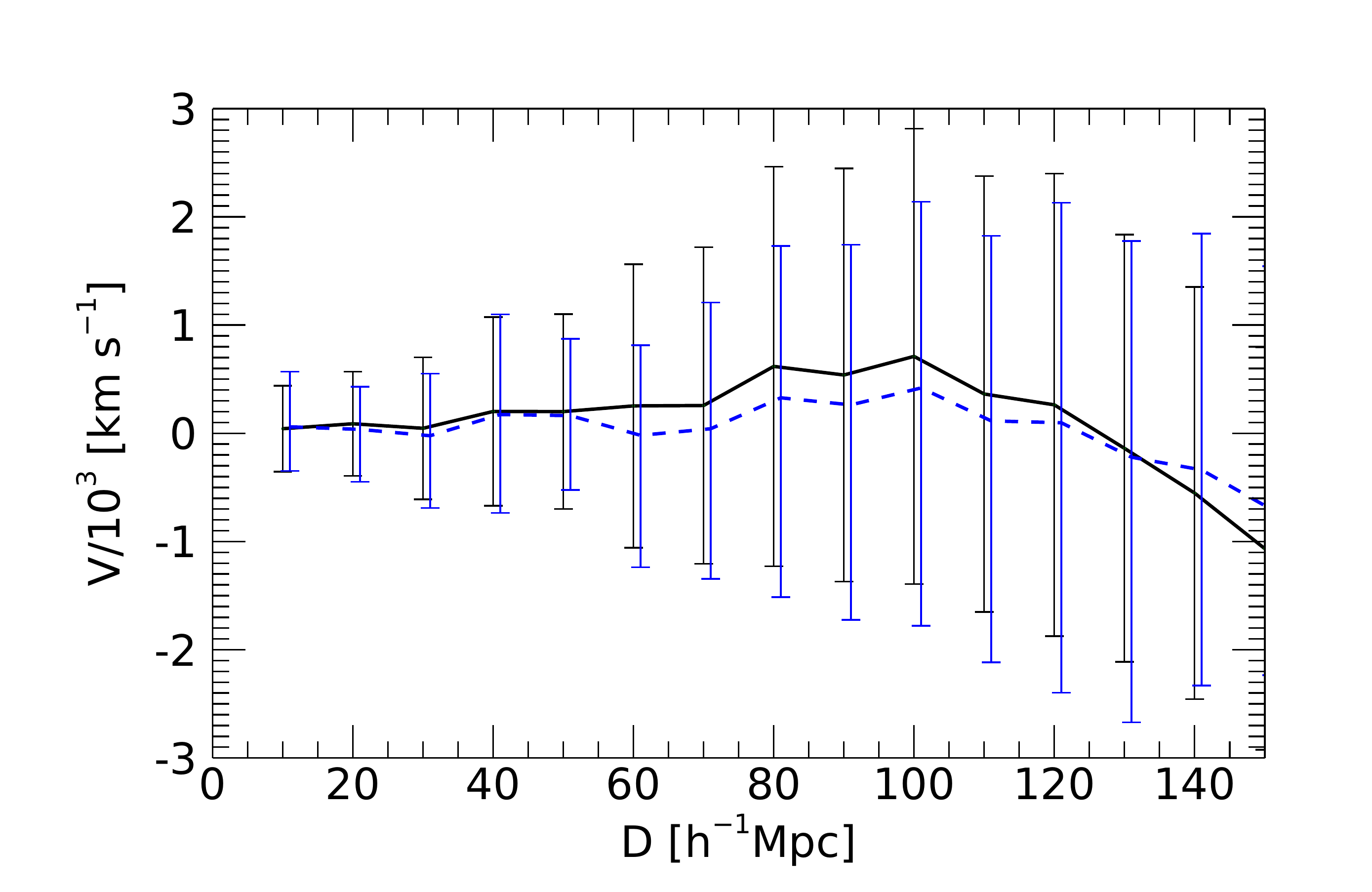}
}
\centerline{
\includegraphics[width=1.0\columnwidth]{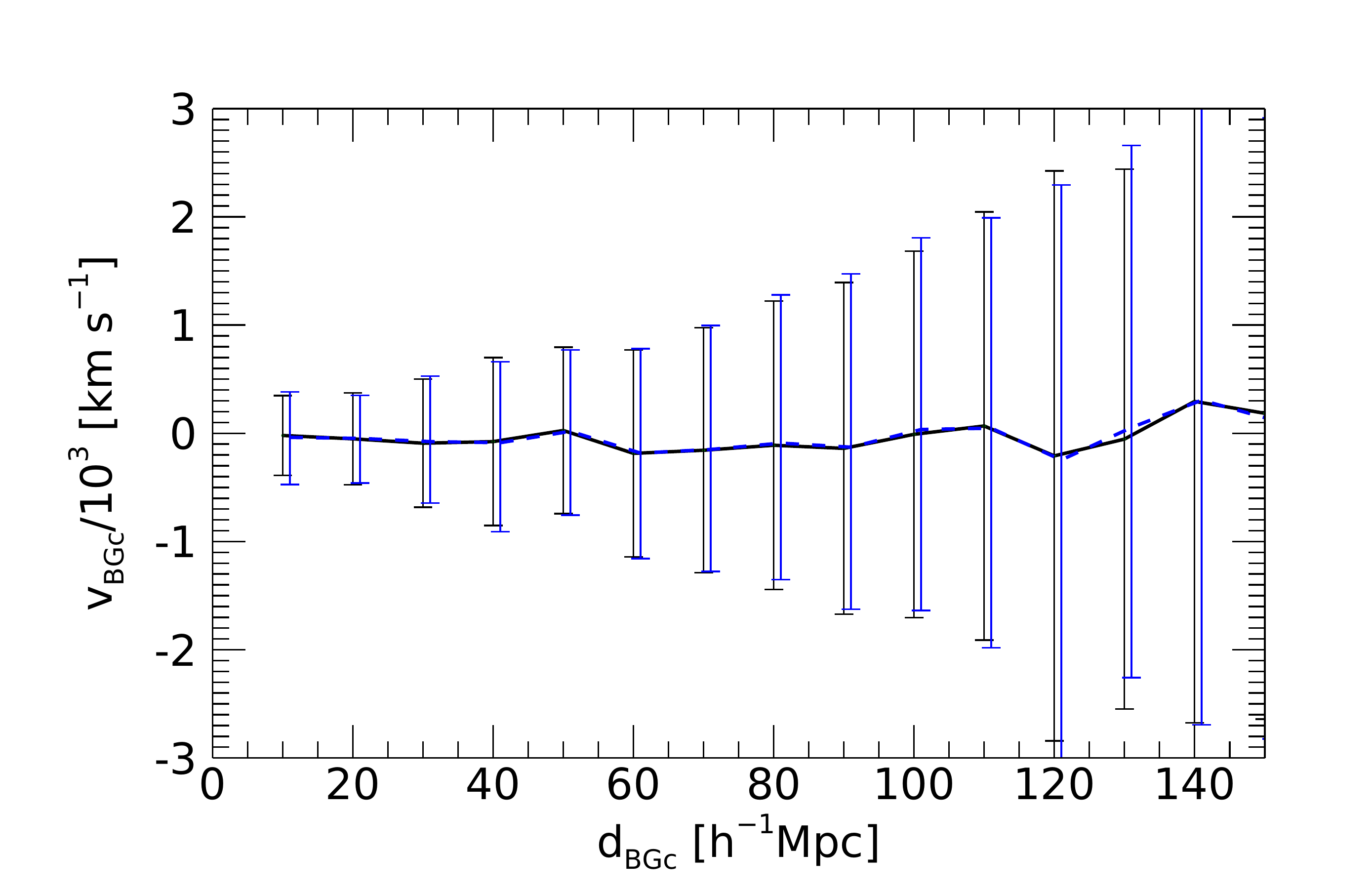}
}
\caption{ 
Mean (black, solid line) and median (blue, dashed line) 
of the   CF3 observed  velocity ($V$)    vs the observed distance ($D$, upper panel) and   of the estimated velocity ($v_{BGc}$)   vs the estimated distance ($d_{BGc}$, lower panel). (Same notations as in Fig. \ref{fig:dBGc-D-residual-dz} .)
} 
\label{fig:Vrad-CF3} 
\end{figure}

Fig  \ref{fig:Vrad-CF3} presents the CF3 data. Its upper panel shows the distribution of  the observed peculiar  radial velocities  vs the observed  distances and its lower panel presents that distribution for the  BGc estimators.  The  bias and the non-Gaussianity of the distribution of the observed velocities are clearly manifested. These are eliminated  by the BGc estimators, the former by the gross deviation of the mean of the velocities from the expected null value and the latter from significant difference between the mean and the median of the velocities in the radial bins.

The estimated Hubble constant from the CF3 data is $H_0 = 75.8 \pm 0.1 \pm 1.0 \pm 0.4 \kmsMpc$,  where the first uncertainty ($0.1$) corresponds to the formal statistical error, the 2nd ($1.0$) to the cosmic variance and the 3rd ($0.4 \kmsMpc$) to the error variance. The estimation of the uncertainties due to the cosmic and error variances are taken from the analysis of the mock catalogs. \cite{2016AJ....152...50T} estimation, based on the analysis of  the peculiar velocities monopole term, is  $H_0 = 75  \pm 2   \kmsMpc$.

\section{Discussion}
\label{sec:disc}

Surveys of   peculiar velocities of galaxies are actually surveys of their luminosity distances and redshifts and  the proper distances  and  peculiar velocities are derived quantities. Within that context the redshifts are (relatively) easily obtained and are accurately known. Estimating  distances  from observed astronomical data is challenging. Going beyond  the formidable challenge of the zero-point calibration  of the various distance indicator one faces the issue that the observational 
 errors are  Gaussian distributed on the distance moduli and thereby are lognormal on the actual distances and on the peculiar velocities. This  is  the lognormal bias. A scheme, the bias Gaussinization correction (BGc),  is presented here to undo that bias. This is done by  constructing a subsample of data points at a given redshift and Gaussianizing the distribution of their distances and of their peculiar velocities, while retaining the medians  of these  distribution. 
 Consequently,  the basic relation between the Hubble velocity, hence distance, peculiar velocity and the redshift holds statistically, in the sense of mean values, per the given subsample.
 It is the invariance of the median  under  the Gaussianization transformation which  preserves the underlying monopole moment of the velocity field and suppresses the lognormal bias.

 Some of the robust conclusions to be drawn from the paper are: a. The BGC algorithm eliminates the lognormal bias; b. In the context of the WF/CRs reconstruction the BGc corrected data well approximates the case of an ideal data made  of  exact distances and  normal errors added to the peculiar velocities; c. The dipole moment, i.e. the bulk velocity, of the WF reconstructed velocity field from the BGc data closely recovers the dipole moment of the underlying velocity field all the way to the edge of the data.; d. The monopole moment, namely the mean of appropriately scaled divergence of the velocity field, of the WF/BGc reconstruction  is a good proxy to the true monopole    from $R\ \sim\ 70\ \hmpc$ to the edge of the data; e. The error variance of the monopole and dipole moments is  negligible compared with  the cosmic variance; f. The mean and variance of the monopole and dipole moments of the WF/BGc reconstructed velocity field are in a good agreement  with the mean and cosmic variance derived from the numerical simulation.

Our main motivation for the introduction of the BGc  algorithm  has been the construction of  CF3 data of unbiased distances  and velocities - one that can be used for cosmological parameters estimation and in particular for the the reconstruction of the large  scale structure of the nearby universe. i.e. the local density and 3D velocity fields.
More specifically, we looked for an unbiased estimator of the CF3 data so as  to use  it for an unbiased WF/CRs reconstruction of the local universe. The WF/CRs reconstruction from the BGc corrected CF3 data is to be presented elsewhere (Hoffman et al, in prep).

The BGc handles only issues related to the lognormal bias. It does not address potential biases and uncertainties associated with the zero-point calibration of the different data sets that make the CF3 data. No  attempt is made here to address issues related to the zero-point calibration and it is taken at its face value.

The main caveat  to be concerned with the BGc is  that it acts to correct the 1-point PDF of the velocities without  any reference to the 2-point distribution function. 
This can be justified by the fact that the amplitudes of the observed velocities,   apart from the very nearby data points, are heavily dominated by the uncorrelated distance errors.   The affect of that caveat is investigated in our  upcoming WF/CRs paper.

Of particular interest is the comparison of the BGc treatment of the bias with that of the Bayesian MCMC method of \cite{2019MNRAS.488.5438G}. Both methods have  been applied to the CF3 mock catalogs and are compared by means of their reconstructed density and velocity fields (Valade et al,  in prep). Preliminary results indicate  an overall good agreement between the two - they both recover the `same (mock) universe'. Yet, some subtle differences exist and they are currently   being  explored. 

\section*{Data Availability Statement}
The modifications to observed {\it Cosmicflows-3} distances resulting from the BGc analysis are compiled in a file hosted at the {\it Extragalactic Distance Database} (http://edd.ifa.hawaii.edu), in the section {\it Summary Distances}, labeled {\it CF3 Modelled Group Distances}.  The file includes BGc distances and inferred peculiar velocities and, for comparison, the equivalent information from the \cite{2019MNRAS.488.5438G} Bayesian MCMC analysis and the directly observed values.
\section*{Acknowledgements}
This work has been done within the framework of the Constrained Local UniversE Simulations (CLUES) simulations.
YH has been partially supported by the Israel Science Foundation grant ISF 1358/18.
AN has been supported by the Israel Science Foundation grant 
ISF 936/18.
NIL \& AV  acknowledge financial support of the Project IDEXLYON at the University of Lyon under the Investments for the Future Program (ANR-16-IDEX-0005). RBT was supported in the development of the Cosmicflows-3 collection of distances by US National Science Foundation award AST09-08846 and NASA award NNX12AE70G.

\bibliographystyle{mn2e}
\bibliography{BGc20}

\bsp
\label{lastpage}
\end{document}